\documentclass[3p]{elsarticle}

\usepackage{amssymb}
\usepackage[figuresright]{rotating}
\usepackage{subfigure}
\usepackage{xurl}
\usepackage[hidelinks,colorlinks=true,linkcolor=blue,citecolor=blue]{hyperref}
\usepackage[dvipsnames]{xcolor}
\usepackage{lineno}
\begin{document}

\begin{frontmatter}

%% Title, authors and addresses

%% use the tnoteref command within \title for footnotes;
%% use the tnotetext command for the associated footnote;
%% use the fnref command within \author or \address for footnotes;
%% use the fntext command for the associated footnote;
%% use the corref command within \author for corresponding author footnotes;
%% use the cortext command for the associated footnote;
%% use the ead command for the email address,
%% and the form \ead[url] for the home page:
%%
%% \title{Title\tnoteref{label1}}
%% \tnotetext[label1]{}
%% \author{Name\corref{cor1}\fnref{label2}}
%% \ead{email address}
%% \ead[url]{home page}
%% \fntext[label2]{}
%% \cortext[cor1]{}
%% \address{Address\fnref{label3}}
%% \fntext[label3]{}

%\dochead{}
%% Use \dochead if there is an article header, e.g. \dochead{Short communication}
%% \dochead can also be used to include a conference title, if directed by the editors
%% e.g. \dochead{17th International Conference on Dynamical Processes in Excited States of Solids}

\title{Structured Multifractal Scaling of the Principal Cryptocurrencies:
Examination using a Self-Explainable Machine Learning}

\author[1,2,3]{Foued SA\^{A}DAOUI\corref{mycorrespondingauthor}}
\cortext[mycorrespondingauthor]{Corresponding author}
\ead{fsadawi@kau.edu.sa; foued.saadaoui@isgs.rnu.tn}

%\author[2,4]{Othman BEN MESSAOUD}
%\ead{bmo.stat@gmail.com}

\address[1]{Department of Statistics, Faculty of Sciences, King Abdulaziz University, P.O BOX 80203, Jeddah 21589, SAUDI ARABIA}

\address[2]{Lab: LR18ES15 Algebra, Number Theory \& Nonlinear Analysis, Faculty of Sciences, Monastir 5019, TUNISIA}
\address[3]{University of Sousse, Institut des Hautes Etudes Commerciales (IHEC), Sousse 4054, TUNISIA}

\address[4]{University of Tunis, Institut Sup\'{e}rieur de Gestion (ISG) de Tunis, Cit\'{e} Bouchoucha, Tunis 2000, TUNISIA}

\begin{abstract}

%
%Multifractal analysis is a mathematical tool that provides a
%robust characterization of data in terms of pointwise regularity
%properties.
%However, for time series with structural changes, the use of the
%best-known methods, such as multifractal detrended fluctuation
%analysis (MF-DFA) and its asymmetric and skewed variants, does not
%take into account the effect of multiple breaks and regime
%changes. This, therefore, risks biasing the statistics of these
%tests.
Multifractal analysis is a forecasting technique used to study the
scaling regularity properties of financial returns, to analyze the
long-term memory and predictability of financial markets. In this
paper, we propose a novel structural detrended multifractal
fluctuation analysis (S-MF-DFA) to investigate the efficiency of
the main cryptocurrencies. The new methodology generalizes the
conventional approach by allowing it to proceed on the different
fluctuation regimes previously determined using a change-points
detection test. In this framework, the characterization of the
various exogenous factors influencing the scaling behavior is
performed on the basis of a single-factor model, thus creating a
kind of self-explainable machine learning for price forecasting.
The proposal is tested on the daily data of the three among the
main cryptocurrencies in order to examine whether the digital
market has experienced upheavals in recent years and whether this
has in some ways led to a structured multifractal behavior. The
sampled period ranges from April 2017 to December 2022. We
especially detect common periods of local scaling for the three
prices with a decreasing multifractality after 2018. Complementary
tests on shuffled and surrogate data prove that the distribution,
linear correlation, and nonlinear structure also explain at some
level the structural multifractality. Finally, prediction
experiments based on neural networks fed with multi-fractionally
differentiated data show the interest of this new self-explained
algorithm, thus giving decision-makers and investors the ability
to use it for more accurate and interpretable forecasts.
\end{abstract}
\begin{keyword}
Analytics \sep Change-Point Detection \sep Structural Complexity
\sep Structured Multifractality \sep Cryptocurrencies \sep
Self-Explainable Machine Learning.
%% keywords here, in the form: keyword \sep keyword

%% PACS codes here, in the form: \PACS code \sep code

%% or \MSC[2008] code \sep code (2000 is the default)
\end{keyword}

\end{frontmatter}

\section{Introduction}\label{sec:0}

For most business economists, forecasting cryptocurrency prices
\cite{[Catania19],[Ftiti21],[Guo21]} is a challenging process due
to the highly volatile nature of the market. Unlike traditional
assets, such as stocks, there are no underlying fundamentals, such
as earnings or dividends, that can be used to predict
cryptocurrency prices. This makes it difficult to use traditional
financial analysis techniques. However, nonlinear time series
analysis \cite{[Bouteska23],[Oh10],[Yolcu13]} can be used as a
decision support system to make accurate predictions about
cryptocurrency prices. Some possible methods include. technical
analysis, using chart patterns and indicators to identify
nonlinear trends and make predictions about future price
movements; mathematical models, using historical cryptocurrency
price data to build dynamical models that can predict future
prices; and machine learning, using a variety of algorithmic
techniques, such as neural networks, to analyze historical
cryptocurrency data and make predictions about future prices
\cite{[Leigh02]}. Multifractal models
\cite{[Benmabrouk08],[Mandelb67]}, in particular, are a type of
time series analysis that can capture the complexity and
non-stationarity of financial markets. They can be used to
identify patterns and trends in the data that are not visible
using traditional methods. These models are also important in
finance because they can capture the persistence and long-term
dependence in financial time series data, meaning that the
correlation between observations decays slowly as the lag between
them increases. This is in contrast to short-memory models, which
assume that the correlation decays quickly.

Multifractal models are considered important in economics and
management because they allow accounting for complexity and
fluctuations at different time scales in the historical data. They
can also help investors and policymakers better understand market
movements and make more informed investment and risk management
decisions. In addition, these models provide a better
understanding of the statistical properties of financial time
series, especially the properties of volatility, which is one of
the key factors for investors and traders. Multifractal models
take into account time-varying properties of volatility that are
important for investment decisions. The use of multifractal models
in economics/finance dates back to the 1960s with the illustration
of Beno\^{i}t Mandelbrot's fundamental work on the self-similarity
of U.S. commodity markets in relation to cotton futures
\cite{[Mandelb67]}. A few years later, and especially with the
development of computing and storage technology, interest in
multifractal models has spread even more, especially in the areas
of financial modeling. Several articles on the self-similarity of
stock index returns, exchange rates and energy prices have been
published \cite{[ASL21],[MAE21],[SAA18],[SAA23],[Wala21],[XU21]}.
This stream of work on multifractal modeling of financial data
continued into the cryptocurrency era of the past decade. A
multitude of works have thus been carried out, the results of
which were somewhat in line with previous works on conventional
money markets
\cite{[ASL2-20],[ASSAF23],[Ghaza20],[Mensi19],[MNF20],[Yi22]}.

Several recent papers have addressed the topic of Bitcoin price
multifractality by considering univariate time series. Shresth
\cite{[Shrestha21]} showed that Bitcoin returns are multifractal
and inefficient. By performing further tests, the author also
found that multifractal scaling and inefficiency are basically
caused by autocorrelated returns as well as extreme returns. Using
high-frequency returns of Bitcoin prices, Takaishi
\cite{[Takais18]} investigated the descriptive statistics and
multifractality of Bitcoin. The author found that the distribution
was fat-tailed and that the kurtosis deviated significantly from
that of a Gaussian. By performing a multifractal analysis, he also
proved that the Bitcoin series exhibited self-similarity and that
the autocorrelation and the fat-tailed distribution contributed
significantly to this. His findings also showed that Brexit has
influenced the GBP-USD exchange rate but not Bitcoin. Stosic et
al. \cite{[Stosic19]} explored the multifractal behavior of price
and volume changes of fifty cryptocurrencies using the
Multifractal Detrended Fluctuation Analysis (MF-DFA). Their
statistics indicated that prices were relatively more complex than
volumes, and that large and small fluctuations dominated the
multifractal behavior of price and volume changes, respectively.
The authors also found that there were no autocorrelations in
price changes, while volume changes exhibited anti-persistent
long-term autocorrelations. They also concluded that the
multifractal behavior of the cryptocurrency market is strikingly
similar to that of stock markets, but differs from that of regular
exchange rates.

Another category of research work has instead focused on the
multifractality aspect of cryptocurrencies in relationship with
other influencing factors, such as stock markets, oil and gold.
Zhang et al. \cite{[Zhang19]} studied multifractal
cross-correlations between Bitcoin prices and other financial
markets (gold and USDX). They found that Bitcoin prices and
volumes displayed multifractal characteristics and that
heavy-tailed distributions have a significant contribution to
multifractality. The authors also found significant multifractal
cross-correlations between Bitcoin and gold markets. On the other
hand, the cross-correlations between Bitcoin and USDX were only
significant over the long term. Ghazani and Khosravi
\cite{[Ghaza20]} investigated the cross-correlations between three
benchmark cryptocurrencies (including Bitcoin, Ethereum, and
Ripple) and some of the well-known crude oils (West Texas
Intermediate (WTI) and Brent). In particular, they found the
existence of multifractal cross-correlations across ten bivariate
time series in the study and that the strength of multifractality
between Ethereum and Ripple is the highest, followed by the
strength of multifractality between WTI crude oil and Ethereum.
Telli and Chen \cite{[Telli21]} examined the relationship between
Bitcoin, Ethereum, Litecoin, XRP crypto-markets and public
attention for crypto-assets in terms of changing multifractal
characteristics. Their results showed that the cross-correlations
of the series of social platforms with the returns series have a
different form than those with the changes in volumes.

However, the two points of view discussed above regarding the
multifractal analysis of cryptocurrencies do not take into account
the structural changes occurring after breakpoints ai and their
effect on the multifractality. For this reason, new approaches
such as asymmetric multifractality and skewed multifractality have
come to improve existing techniques and take into account this
scaling variation factor before and after a breakpoint. Telli and
Chen \cite{[Telli20]} for example studied the multifractality of
Bitcoin and gold returns and volatility over full intervals as
well as sub-sampling periods bounded by structural breaks. Using
the MF-DFA approach, they found that Bitcoin returns have a higher
level of multifractality than gold. These results were also
confirmed using a sliding window technique. Kristjanpoller and
Bouri \cite{[Kristj19]} examined long-term cross-correlations and
asymmetric multifractality between currencies like Swiss Franc,
Euro, British Pound, Yen, and Australian Dollar, and major
cryptocurrencies (Bitcoin, Litecoin, Ripple, Monero, and Dash).
The empirical results showed evidence of a significant
cross-correlation asymmetry, which was found to be persistent and
multifractal in most cases. Mensi et al. \cite{[Mensi19]} examined
asymmetric multifractality and weak form efficiency for Bitcoin
and Ethereum. Their results showed evidence of structural breaks
and asymmetric multifractality. Moreover, the multifractality gap
between the uptrend and the downtrend is small when the timescale
is small, but increases as the timescale increases. Finally, Mensi
et al. \cite{[Mensi22]} studied the impact of COVID-19 on price
efficiency and asymmetric multifractality of major financial
assets including Bitcoin. Analysis using a detrended multifractal
asymmetric fluctuation analysis approach (A-MF-DFA) showed
evidence of asymmetric multifractality in all markets that
increase with scales. These findings suggested that the pandemic
intensified the inefficiency of all markets except Bitcoin.

The methodology we propose in this paper is a generalization of
the work of Cao et al. \cite{[Cao13]} and of Sa\^{a}daoui
\cite{[SAA23]} in the sense that it offers a novel technique for
analyzing the multifractal properties not only on two adjacent
intervals but rather on a succession (number greater than or equal
to 2) of adjacent intervals. In this framework, the splitting of
the complete interval is done by proceeding to the preliminary
detection of the various structural change points. A MF-DFA is
thus conducted on each of the resulting sub-samples. We therefore
call this generalized approach structured MF-DFA (S-MF-DFA) since
it allows to determine the levels of local self-similarity on the
sub-samples. %separated by the different structural breaks existing
%in the data set.
These notions essentially based on a single-factor will then allow
us to develop a neural model whose learning sources are mainly
autocorrelations and transitions between regimes (nonlinear
structures). We call this novel algorithm Self-Explained Machine
learning (SX-ML). In this article, the S-MF-DFA approach is
performed on daily data of three among the main cryptocurrencies
(Bitcoin, Ethereum and Litecoin) over a period spanning from April
14, 2017 to December 22, 2022. The cryptocurrency industry is
currently well known as being one of the fastest-growing
industries in the world. Nevertheless, managing its risks remains
one of the major concerns and constraints facing investors. The
sampling period we choose for our experiments is known to be one
of the tensest periods of the last fifty years as it includes
large-scale political-economic upheavals such as Brexit, COVID-19
and the Russian-Russian conflict
\cite{[BenAmeur22],[Ftiti21],[Yousaf22]}. We therefore set the
objective of testing whether large upheavals really affect the
level of local irregularity before and after each breaking point.
The empirical results seem to confirm this hypothesis for the
three cryptocurrencies studied. These findings are also confirmed
by the prediction experiments carried out using the SX-ML It is
therefore important to build any risk management strategy on each
sub-interval independently of the others.

The structure of the article is as follows. A technical background
on the skewed multifractal detrended fluctuation analysis is put
forth in section \ref{SECT1}. The experimental results are
detailed and discussed in Section \ref{SECT3} and the final
section concludes.

\section{Structural Multifractal Analysis}\label{SECT1}

The structural multifractal analysis that we propose is
essentially based on a break detection test, which is firstly
performed to the time series. Once the hypothesis of the
significance of a break is confirmed, a multifractal analysis is
not performed on each of the subintervals. In what follows, we
will first define the change point test, then the MF-DFA approach
applied to each portion of the subdivided signals.

\subsection{Testing for Change-Points}\label{Sub-SECT1}

A change-point can be defined as a sample or time-instant at which
the statistical characteristics of a time series changes
significantly \cite{[BER11],[KIL12],[LVL05]}. The characteristic
in question can be the mean of the time series, its standard
deviation, or a the scaling property, among others. Given a time
series $X=(x_{1}, x_{2},\ldots, x_{N})^{\top}$, and the following
respective subsequence mean and variance functions:
\begin{equation}\label{EQN:ChP1}
\mu(x_{m},\ldots, x_{n})=\frac{1}{n-m+1}\sum_{s=m}^{n}x_{s}
\end{equation}
\begin{equation}\label{EQN:ChP2}
\sigma^{2}(x_{m},\ldots,
x_{n})=\frac{1}{n-m+1}\sum_{s=m}^{n}\{x_{s}-\mu(x_{m},\ldots,
x_{n})\}^{2}.
\end{equation}
The change-point test aims at finding the value of $h$ such that
the following statistic is the smallest:
\begin{equation}\label{EQN:ChP3}
\mathcal{D}_{\mu,\sigma}(h)=(h-1)\sigma^{2}(x_{1},\ldots,
x_{h-1})+(N-h+1)\sigma^{2}(x_{h},\ldots, x_{N}).
\end{equation}
This result can be generalized to also incorporate other
statistics. In this case, the tests finds $h$ minimizing,
\begin{equation}\label{EQN:ChP4}
\mathcal{D}_{\Delta,\chi}(h)=\sum_{i=1}^{h-1}\Delta(x_{i};
\chi(x_{1},\ldots, x_{h-1}))+\sum_{i=h}^{N}\Delta\{x_{i};
\chi(x_{1},\ldots, x_{h-1})\},
\end{equation}
where $\chi(.)$ represents the section empirical estimate, while
$\Delta(.)$ is a deviation measurement. For economic and financial
time series, however, we often have multiple change points. The
generalized detection technique is simple when the number of
breaks is known in advance. On the other hand, when this number is
unknown, what must be done is to add a penalty term to the
residual deviation. This is because adding change points always
decreases the residual error and leads to overfitting. In the
extreme case, each point becomes a breakpoint and the residual
error disappears. If there is a number $H$ of changepoints to
locate, then the function must minimize the following quantity:
\begin{equation}\label{EQN:JH}
\mathcal{D}_{\Delta,\chi}(H)=\sum_{s=1}^{H-1}\sum_{i=h_{s}}^{h_{s+1}-1}\Delta(x_{i};
\chi(x_{h_{s}},\ldots, x_{h_{s+1}-1}))+\theta H,
\end{equation}
where $h_{0}$ and $h_{H}$ are the first and last samples of the
time series, respectively, while $\beta$ is a fixed penalty added
for each change-point.

Once the number of breaks and their localizations are detrained,
we get a set of adjacent time series: $X_{h}$, $h=1,\ldots,H+1$.
Now, the multifractal analysis will be conducted on each of these
subintervals. Rejecting the self-similarity hypothesis of the same
order on the different subsamples means that our time series
exhibits structural multifractal scaling. In the following
subsection, we revisit the best-known multifractal analysis,
MF-DFA, which is independently applied to the resulted subsamples.
It is notable that the above strategy could be improved to take
into account the false alarms of structural change often wrongly
detected, particularly in finance \cite{[BER11],[ELM14]}. The
detection of multiple change points before measuring the
multifractal scaling can also be done using other techniques such
the unsupervised machine learning. It is possible for example to
apply a cluster analysis calibrated by the
Expectation-Maximization algorithm as in
\cite{[Rabbch17],[Saadaoui08],[Sa3d09]}. This avenue could also be
one of the potential extensions of this work.

\subsection{Structural Multifractal Detrended Fluctuation Analysis}\label{SECT2}

We consider the discrete-time stochastic process
$\{x(t)\}_{t\in\mathbb{N}}$ of length $T$, and define its
fluctuation function as:
\begin{equation}\label{eqn:0}
    f_{t}=|\log_{10}\{x(t+1)/x(t)\}|,~~ t=1,\ldots,T.
\end{equation}
The transformation $f_{t}$ is then split into $H$ adjacent subsets
$f_{h,t}$, $h=1,\ldots,H$ of lengths $T_{h}$, verifying
$f_{t}\equiv\bigcup_{h}f_{h,t}$ and
$\bigcap_{h}f_{h,t}\equiv\varnothing$. The profile statistics of
the subsets $f_{h,t}$ are determined as follows:
\begin{equation}\label{eqn:1}
Y^{(h)}(j)= \sum_{t=1}^{j} \{f_{h,t}- \bar{f}_{h}\},~~
j=1,\ldots,T_{h},
\end{equation}
where $\bar{f}_{h}$ is the average of $f_{h,t}$ over the whole
$h$th segment. Each $Y^{(h)}(j)$ is then split into $T_{h,s}\equiv
\int(T_{h}/s)$ equal non-overlapping segments of length $s$. Since
the length of each subset $T_{h}$ is often not multiple of $s$, a
short part at the end of the profile often rests. Therefore,
another splitting procedure is redone from other side. This gives
overall a number of $N_{s}=2\times T_{h,s}$ segments.

We now estimate the local-trend for each of the $N_{s}$ portions
by a smoothing of the time series piece. Dispersions over each one
of the smooths are expressed as:
\begin{equation}\label{eqn:2}
\sigma_{h}^{2}(\gamma,s) \equiv \frac{1}{s} \sum_{k=1}^{s} \{
Y^{(h)}[(\gamma - 1)s+k]
 -Q_{\gamma}^{(m)}(k)\}^{2},~~~\hbox{for}~\gamma=1,\ldots,T_{h,s}
\end{equation}
and
\begin{equation}\label{eqn:3}
 \sigma_{h}^{2}(\gamma,s) \equiv \frac{1}{s} \sum_{k=1}^{s} \{Y^{(h)}[T_{h}-(\gamma-T_{h,s})s + k]
 -Q_{\lambda}^{(m)}(k)\}^{2}
\end{equation}
for $\gamma=T_{h,s} +1,\ldots,N_{s}$. The functions
$Q_{\gamma}^{(m)}(j)$ in Eqs (\ref{eqn:2}) and (\ref{eqn:3}) are
$m$th ordered fitting polynomials in $\gamma$'s. Averaging over
all segments within each subset $h$, we get the $q$th-ordered
fluctuation function
\begin{equation}\label{eqn:4}
\varphi^{(h)}_{q}(s)\equiv \biggl\{ \frac{1}{N_{s}}
\sum_{\gamma=1}^{N_{s}}[\sigma_{h}^{2}(\gamma,s)]^{q/2}\biggr\}^{1/q}.
\end{equation}
The objective is to measure how, over the $h$ different levels,
the $q$-dependent fluctuation functions $\varphi^{(h)}_{q}(s)$
behave when varying the time-scale $s$ and the values of $q$. This
necessitates to calculate Eq.(\ref{eqn:4}) for several levels of
$s$. Noteworthy that the conventional detrended fluctuation
analysis is obtained for the special case $q=2$, when no
change-point within the time series is detected, i.e., $H=0$.
Sa\^{a}daoui's asymmetric multifractal approach \cite{[SAA23]} is
also obtained when only one break is detected.

A MF-DFA is performed on each one of the $H$ segments by analyzing
log-log plots of $\varphi^{(h)}_{q}(s)$ versus $s$ for each value
of $q$. When the autocorrelation function of $x_{t}$ decays as a
power-law, $\varphi^{(h)}_{q}(s)$ increases, for large values of
$s$, as a power-law, i.e., we have:
\begin{equation}\label{eqn:5}
 \varphi^{(h)}_{q}(s) \propto s^{\rho_{h}(q)},
\end{equation}
where $\rho_{h}(q)$ is the generalized Hurst parameter. Noteworthy
that the particular case $\rho_{h}(0)$, which is the limit
$\rho_{h}(q)$ when $q\rightarrow 0$, is got by applying a
logarithm average, i.e.,
\begin{equation}\label{eqn:6}
  \varphi^{(h)}_{0}(s) \equiv \exp \biggl \{\frac{1}{2N_{s}} \sum_{\gamma=1}^{N_{s}}\ln{[ \sigma_{h}^{2}(\gamma,s) ]}  \biggr\} \propto
 s^{\rho_{h}(0)}.
\end{equation}
Nevertheless, as explained in \cite{[KAN02]}, MF-DFA can only
determine positive values of the hurst exponent, but also becomes
significantly inaccurate for strongly anti-correlated data when
$\rho_{h}(q)$ is almost null. This forces us to consider the
profile from the original data $x_{t}$ by a double summation.

When the data satisfy $x_{t}\geq 0$ and $\sum_{t=1}^{T}x_{t}=1$, a
simple Fluctuation Analysis (FA) can be performed instead of the
DFA. In this case we have the following:
\begin{equation}\label{eqn:10}
\biggl \{ \frac{1}{N_{s}} \sum_{\gamma=1}^{N_{s}}\mid
Y^{(h)}(\gamma s)-Y^{(h)}((\gamma-1)s)\mid^{q} \biggl \}^{1/q} ~~
\propto ~~ s^{\rho_{h}(q)},
\end{equation}
which, by supposing that the size of a subset $T_{h}$ is a
whole-number multiple of $s$, can be written as
\begin{equation}\label{eqn:11}
\sum_{\gamma=1}^{N_{s}} \mid Y^{(h)}(\gamma
s)-Y^{(h)}((\gamma-1)s)\mid^{q} ~~ \propto
 ~~ s^{q\rho_{h}(q)-1}.
\end{equation}
In the multifractal formalism, the difference quantity within the
absolute value operator is defined as box probability for
standardized series and is commonly denoted as:
\begin{equation}\label{eqn:12}
\mathrm{p}^{(h)}_{s}(\gamma) \equiv \sum_{t=(\gamma-1)
s+1}^{\gamma s} x_{t} = Y^{(h)}(\gamma s) - Y^{(h)}((\gamma-1)s).
\end{equation}

At each $h$th segment, for $q\in \mathbb{R}$, the scaling exponent
$\tau^{(h)}(q)$ can, on the other hand, be expressed via a
partition function $Z_{q}(s)$ with the following relationship:
\begin{equation}\label{eqn:13}
Z^{(h)}_{q}(s) \equiv \sum_{\gamma=1}^{T_{h}/s} \mid
\mathrm{p}^{(h)}_{s}(\gamma) \mid ^{q} ~~ \propto ~~
s^{\tau^{(h)}(q)}.
\end{equation}
From Eqs. (\ref{eqn:11}), (\ref{eqn:12}) and (\ref{eqn:13}) we can
explicitly get the link between the two scaling exponents,
\begin{equation}\label{eqn:14}
\tau^{(h)}(q)= q \rho_{h}(q) -1.
\end{equation}
An alternative method to analyse the multifractal character is by
expressing its singularity spectrum $f(\alpha_{h})$. This one is
linked to $\tau^{(h)}(q)$ via the Legendre transformation, i.e.,
\begin{equation}\label{eqn:15}
\alpha_{h}=\frac{\mathrm{d}\tau^{(h)}(q)}{\mathrm{d}q} ~~
\hbox{and} ~~ f(\alpha_{h})=q \alpha_{h} - \tau^{(h)}(q),
\end{equation}
where $\alpha_{h}$ is the H\"{o}lder exponent, while
$f(\alpha_{h})$ is the dimension of the part of the time series
characterized by $\alpha_{h}$.

Using Eqs (\ref{eqn:14}) and (\ref{eqn:15}), we obtain
\begin{equation}\label{eqn:16}
\alpha_{h}= \rho_{h}(q) + q \rho_{h}'(q) ~~ \hbox{and} ~~
f(\alpha_{h}) = q[\alpha_{h}-\rho_{h}(q)] +1.
\end{equation}
If the $h$th subset of the time series is multifractal, its
generalized Hurst exponent can be expressed as:
\begin{equation}\label{eqn:17}
\rho_{h}(q)=\frac{1}{q}-\frac{\ln(\beta_{1}^{q}+\beta_{2}^{q})}{q\log
2}~~~~(\beta_{1}>\beta_{2}),
\end{equation}
where $\beta_{1}$ and $\beta_{2}$ are the regression parameters.
The multifractal level is equivalent to the range of the spectrum
$\Delta\alpha_{h}=\alpha_{h,\max}-\alpha_{h,\min}$. Finally, we
have a particular case, when $q\rho_{h}'(q)$ is close zero as
$q\rightarrow\pm\infty$, $\Delta\alpha_{h}$ is given by
$\rho_{h}(-\infty)-\rho_{h}(+\infty)=$$\frac{\log(\beta_{2})-\log(\beta_{1})}{\log(2)}$.

\section{Empirical Results}\label{SECT3}

\subsection{Descriptive Statistics and Tests}\label{Sub-SECT2}

\begin{figure}[!ht]
  \centering
  \subfigure[Bitcoin prices]{\label{fig:log-prices}\includegraphics[width=0.45\textwidth]{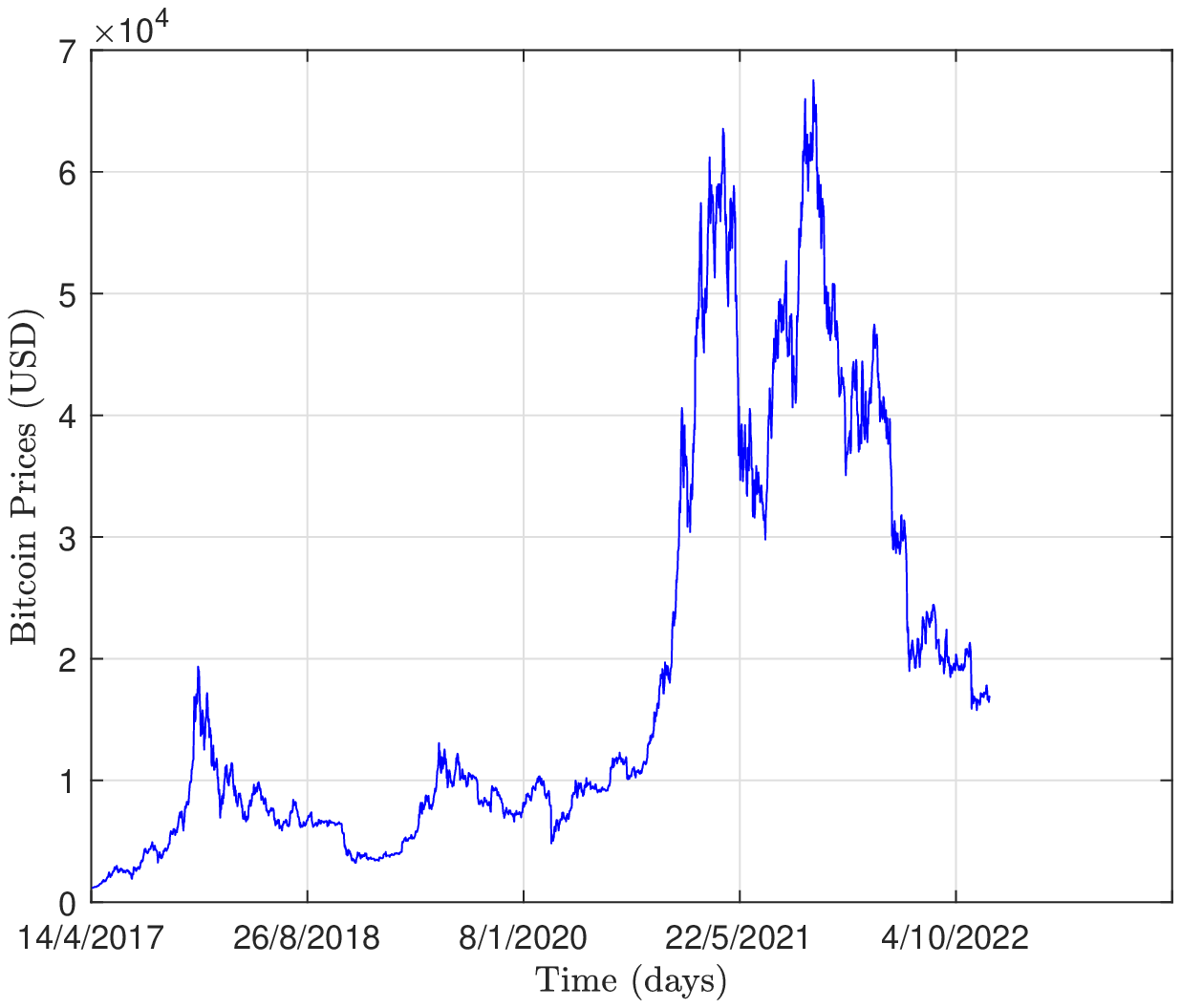}}
  \subfigure[Bitcoin log-prices]{\label{fig:autocorr}\includegraphics[width=0.45\textwidth]{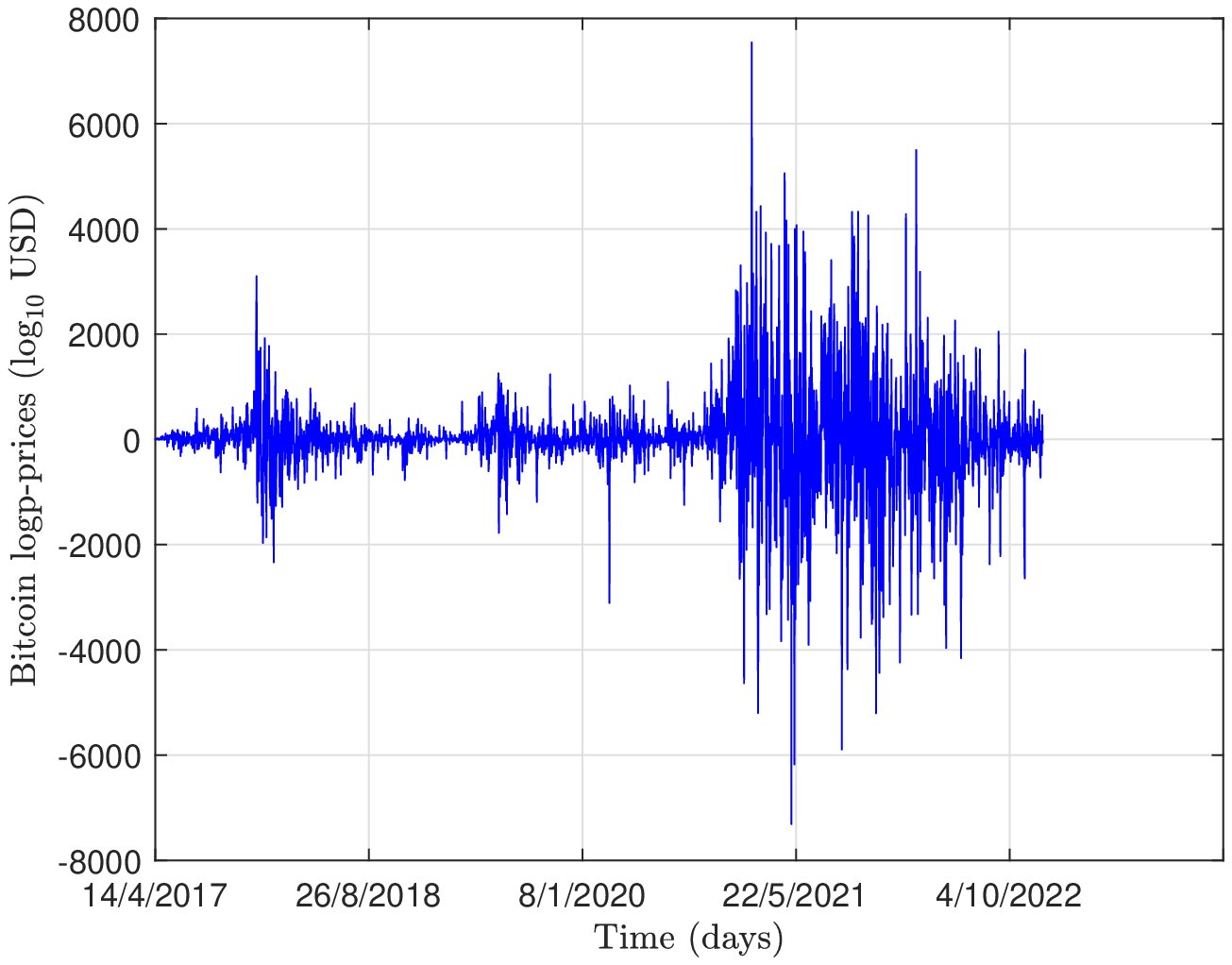}}
  \subfigure[Ethereum prices]{\label{fig:log-prices}\includegraphics[width=0.45\textwidth]{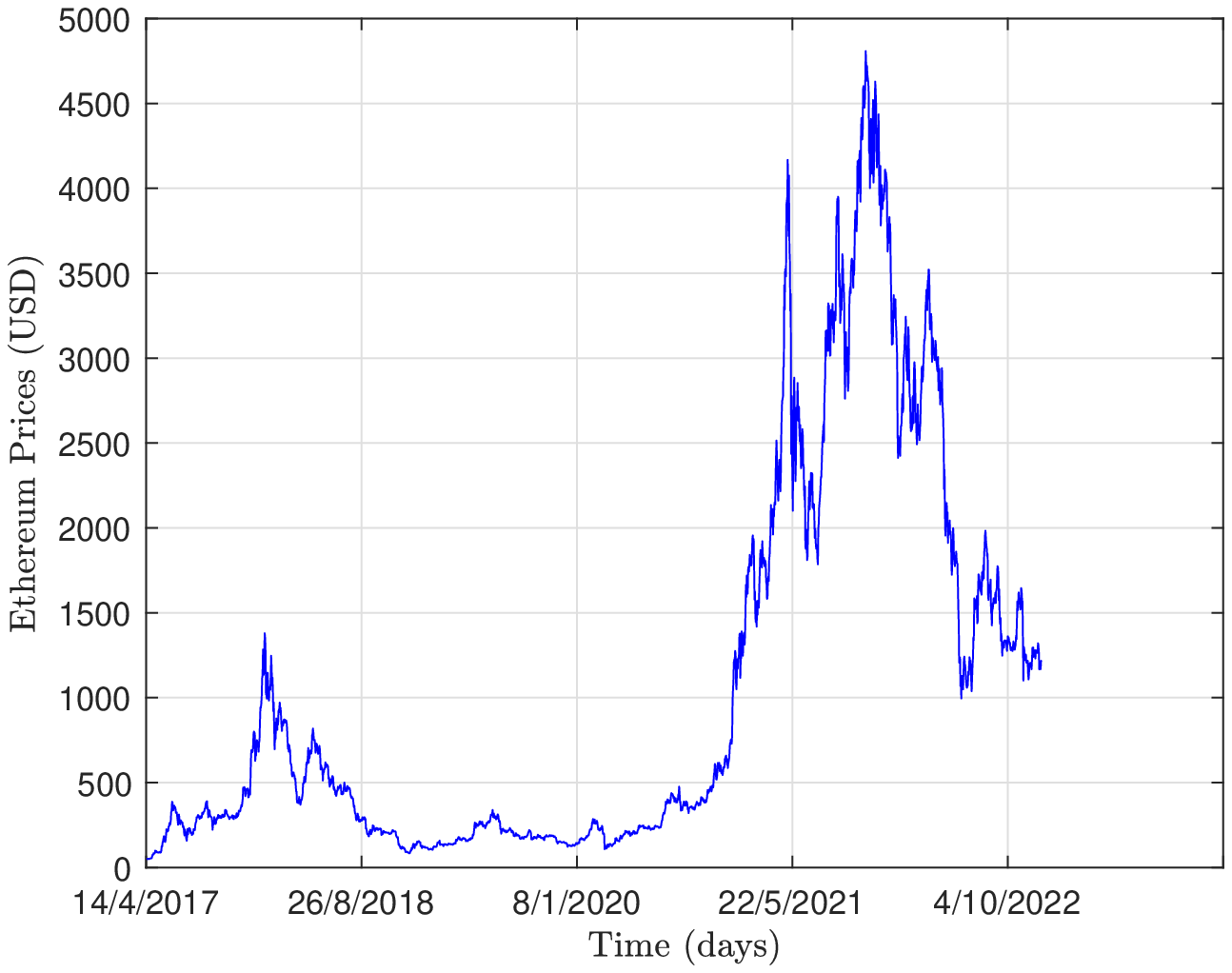}}
  \subfigure[Ethereum log-prices]{\label{fig:autocorr}\includegraphics[width=0.45\textwidth]{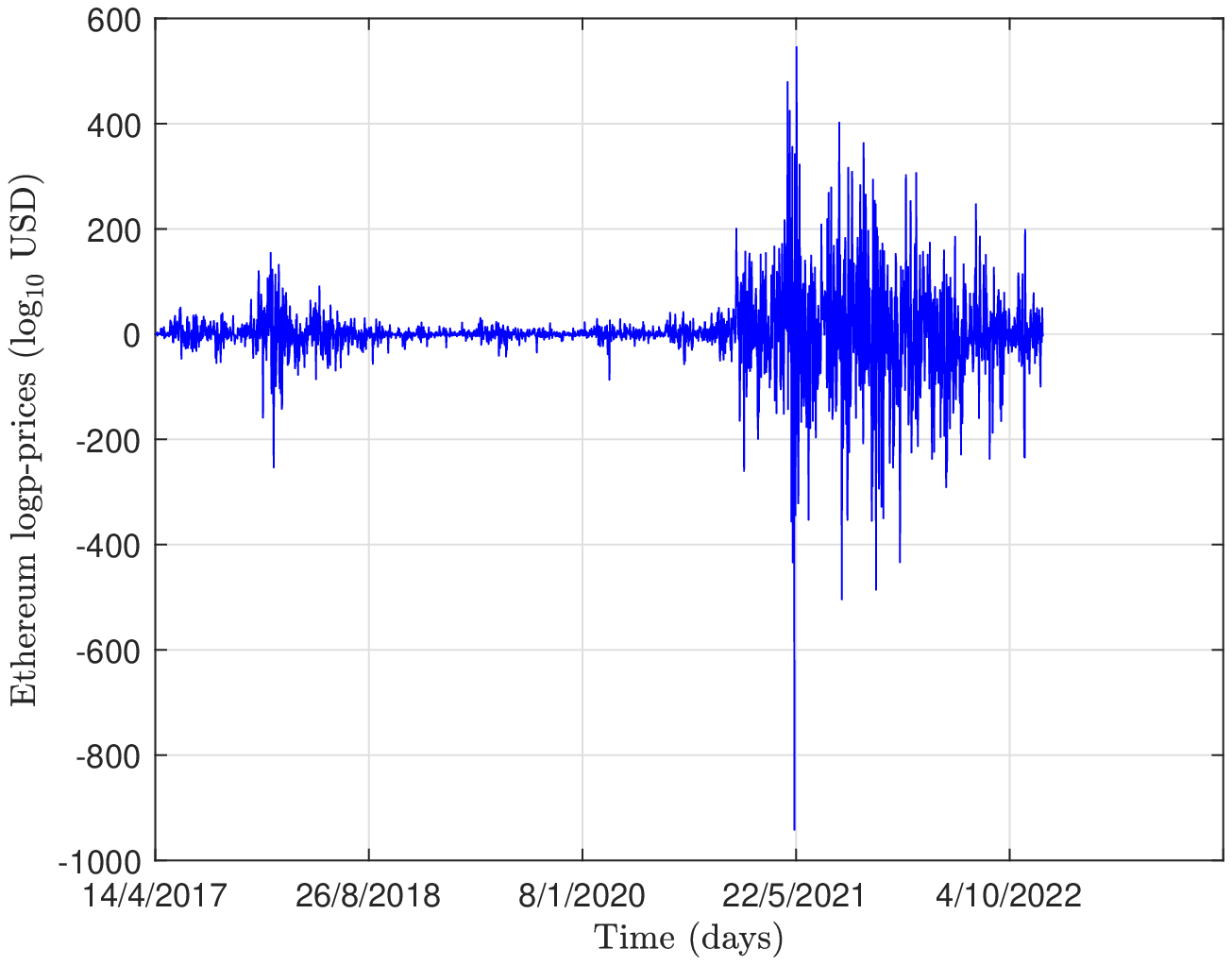}}
  \subfigure[Litecoin prices]{\label{fig:log-prices}\includegraphics[width=0.45\textwidth]{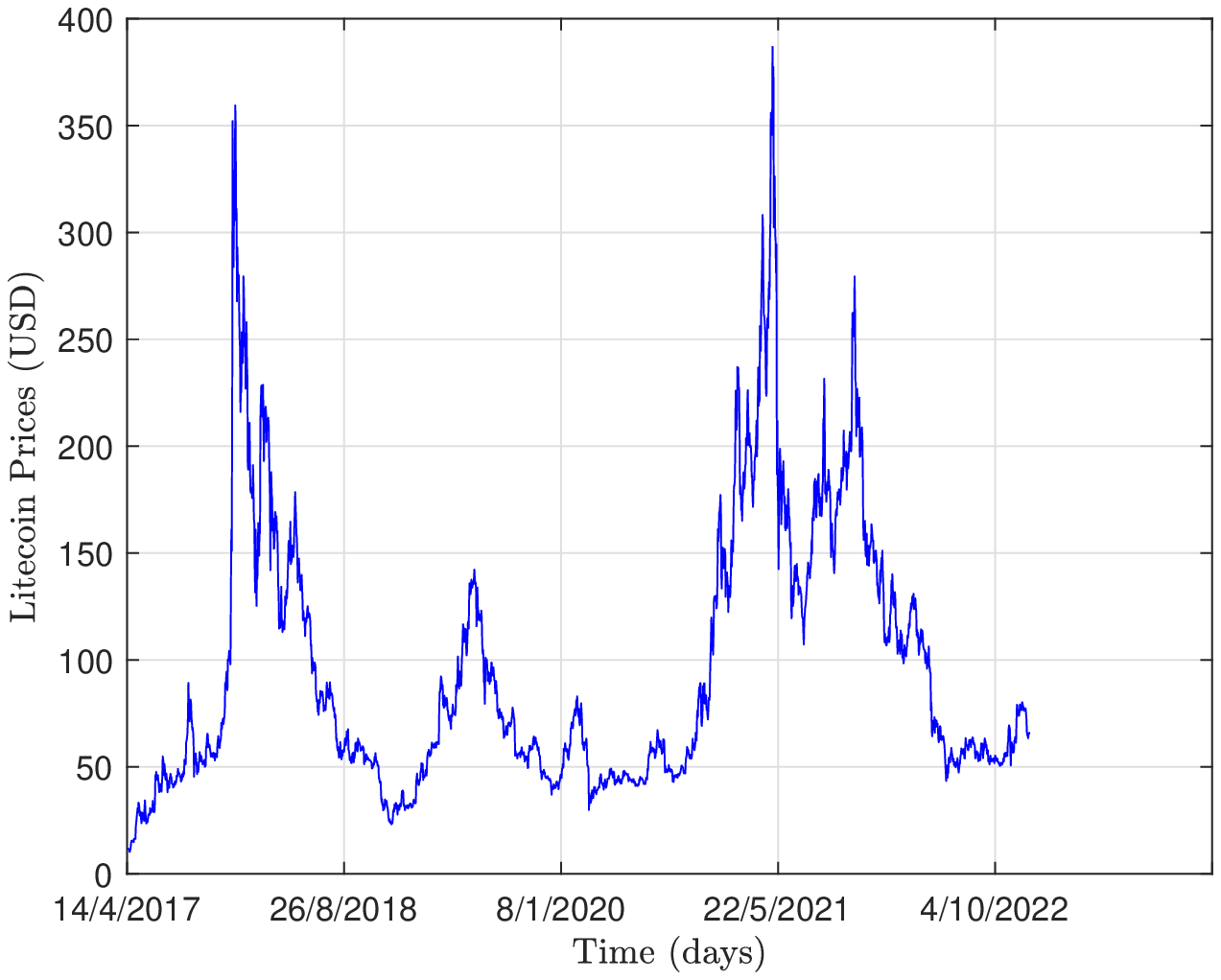}}
  \subfigure[Litecoin log-prices]{\label{fig:autocorr}\includegraphics[width=0.45\textwidth]{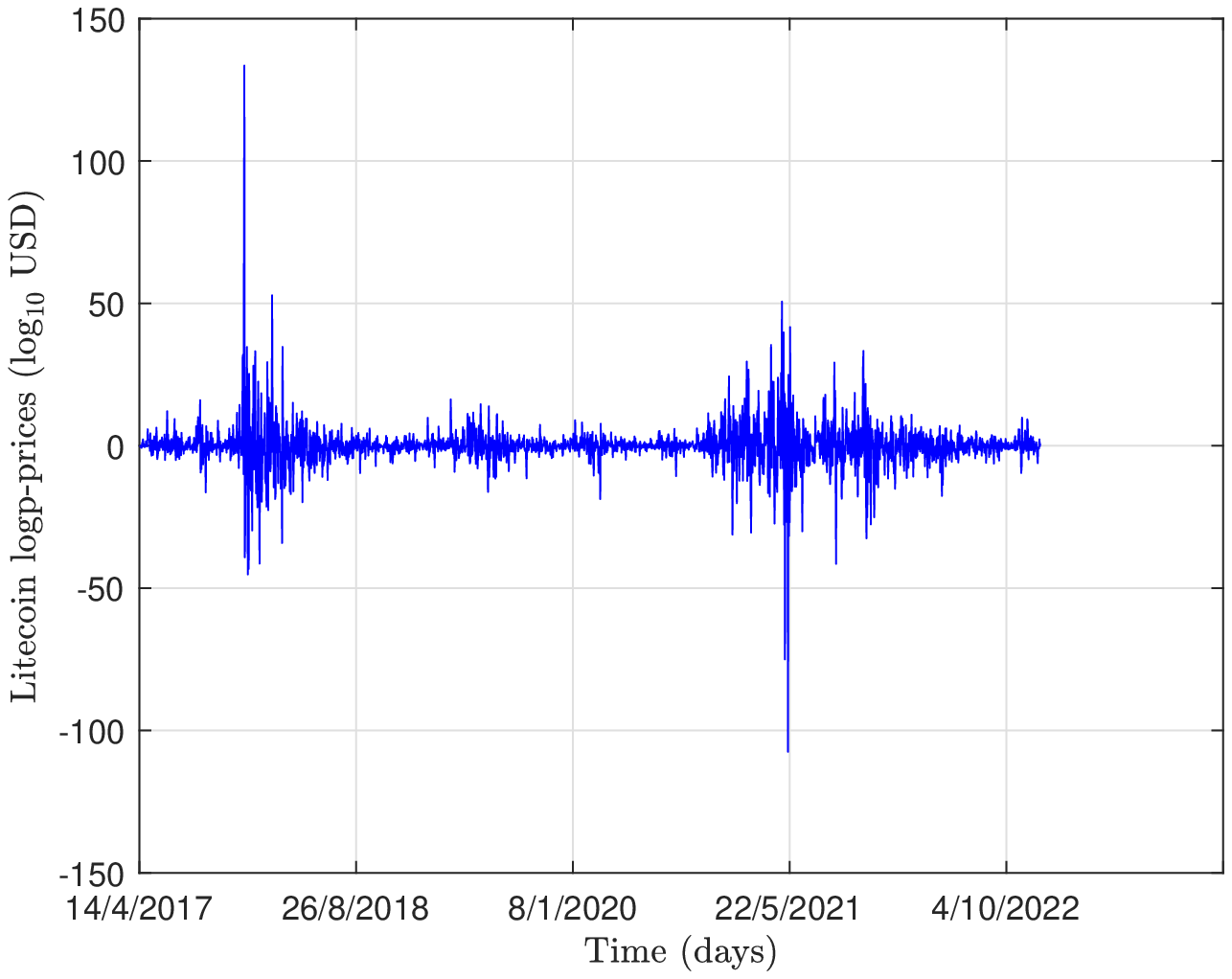}}
  \caption{Price and log-price time series of Bitcoin, Ethereum and Litecoin cryptocurrencies. The samples cover a daily period
ranging from April 14, 2017, to December 22, 2022.}
  \label{fig:plots1}
\end{figure}

The raw time series of Bitcoin, Ethereum and Litecoin prices as
well as their log-prices are firstly plotted in Figure
\ref{fig:plots1}. The samples cover a daily period ranging from
April 14, 2017, to December 22, 2022. This period is known for its
difficult politico-economic conditions including in particular the
global health crisis linked to the Coronavirus in 2019 (COVID-19)
as well as the war in Eastern Europe taking place until the
present time between Russia and the Ukraine. In this subsection,
we will carry out a descriptive statistical analysis as well as a
visualization of the data in order to identify the stylized facts.
We then use the changing point test defined above to detect a
significant change in the time series of the three cryptocurrency
prices. By briefly reading the time series plots in Figure
\ref{fig:plots1}, we can also see some similarity in the curves of
the three cryptocurrencies. The succession of upheavals in recent
years is certainly one of the main causes of this co-movement. In
fact, generally speaking, in periods of crisis in the global
economy, all the indices tend to go parallel in the wrong
direction. Here, we notice the opposite, prices rise significantly
through tense periods. It is very clear, for example, that the
health crisis linked to COVID-19 has led to record levels in the
prices of cryptocurrencies. It is likely that the rise of remote
working during the times of confinement has contributed to this
phenomenon.

The descriptive statistics and goodness-of-fit tests are given in
Tables \ref{Tab:Descriptive} and \ref{Tab:extremes}. We
essentially report the main estimators of central tendencies,
dispersions, outliers, comparisons with the Gaussian distribution
as well as long memory. We can especially notice that the three
prices show almost the same statistical characteristics.
Asymmetry, excess kurtosis, nonlinearity, long memory and the
presence of mild and extreme outliers\footnote{Mild outliers are
observations that are outside the interval $[Q_{1} - 1.5\times
IQR~,~Q_{3} + 1.5\times IQR]$ ($Q_{1}$ and $Q_{3}$ are resp. the
first and third quartiles, while $IQR$ stands for interquartile
range). Extreme outliers are observations that are outside the
interval $[Q_{1} - 3\times IQR~,~ Q_{3} + 3\times IQR$.}, are the
common facts for the three cryptocurrencies. We can particularly
note the presence of a considerable proportion of maximum outliers
while an absence of minimum outliers. This asymmetry in the
distribution of extreme values is also an interesting fact which
deserves to be studied in greater depth. The frequency spectra
(MUSIC: multiple signal classification) and kernel estimates of
the distributions in Figure \ref{fig:plots3.} show that the
spectral and probabilistic densities are relatively similar for
the returns of the three cryptocurrencies. All these statistics
therefore point to a co-movement of the main assets, which in fact
is not the preferred scenario for risk-averse investors, since the
opportunities for diversification in this case are very limited.
It is notable that all results are obtained on MATLAB 9.0
\cite{[Matlab]} under Microsoft Windows 8 as operating system and
on a computer with Intel(R) Core(TM)i5-6400 CPU@2.70GHZ with 16GB
RAM.

\subsection{Testing Structured Multifractality}\label{Sub-SECT2}

\begin{figure}[!ht]
  \centering
  \subfigure[Bitcoin: One change-point]{\label{fig:log-prices}\includegraphics[width=0.45\textwidth]{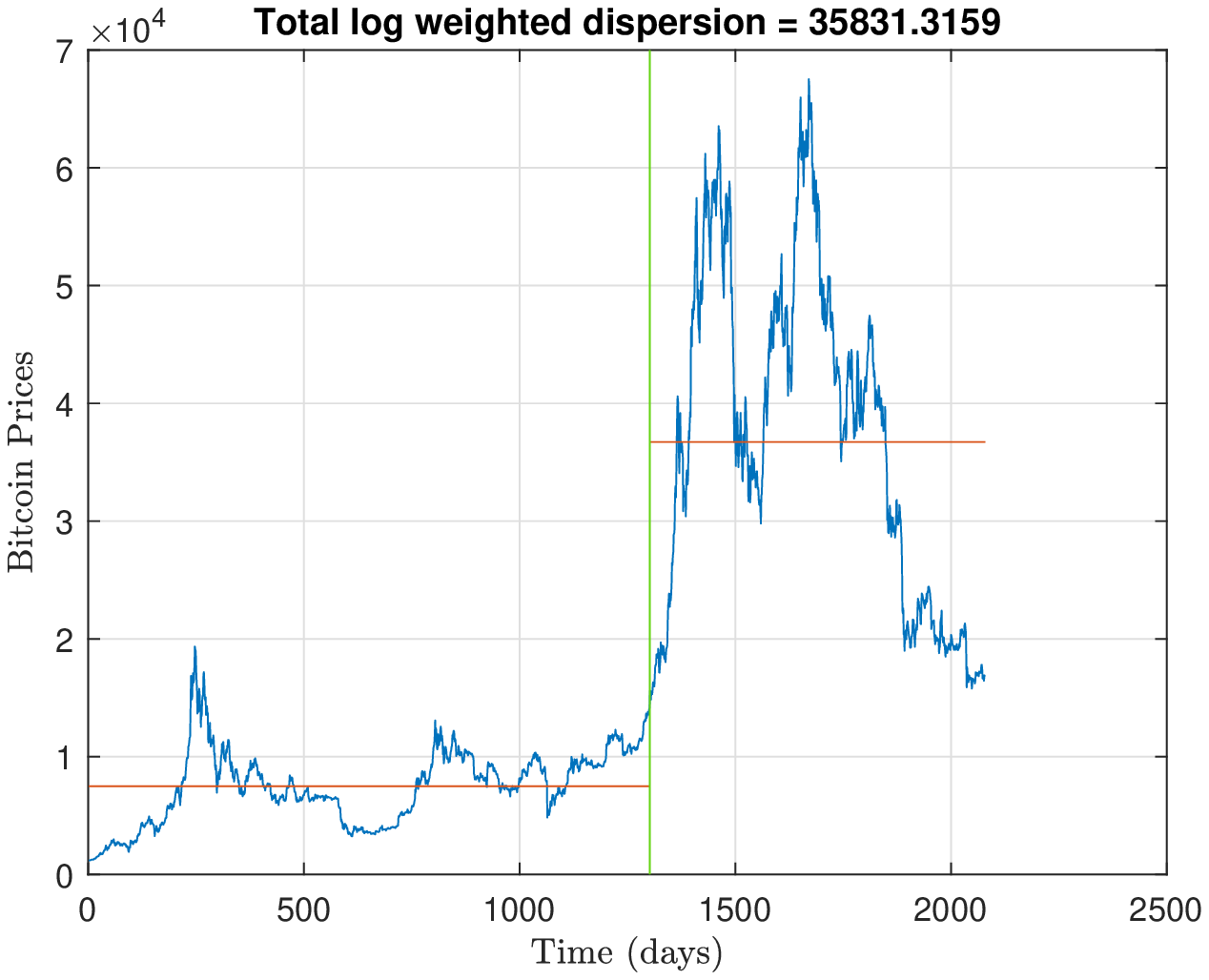}}
  \subfigure[Bitcoin: Four change-points]{\label{fig:autocorr}\includegraphics[width=0.45\textwidth]{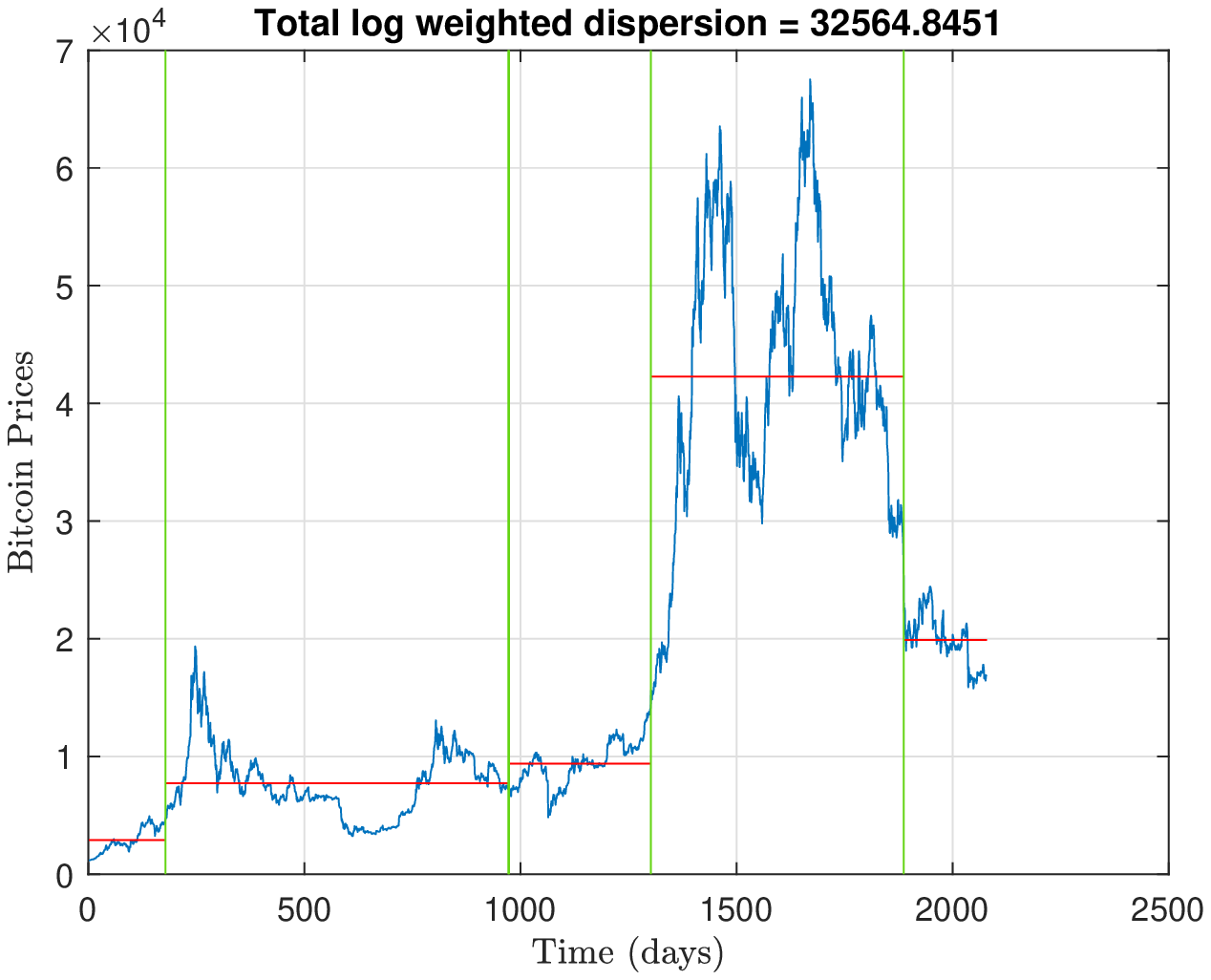}}
  \subfigure[Ethereum: One change-point]{\label{fig:log-prices}\includegraphics[width=0.45\textwidth]{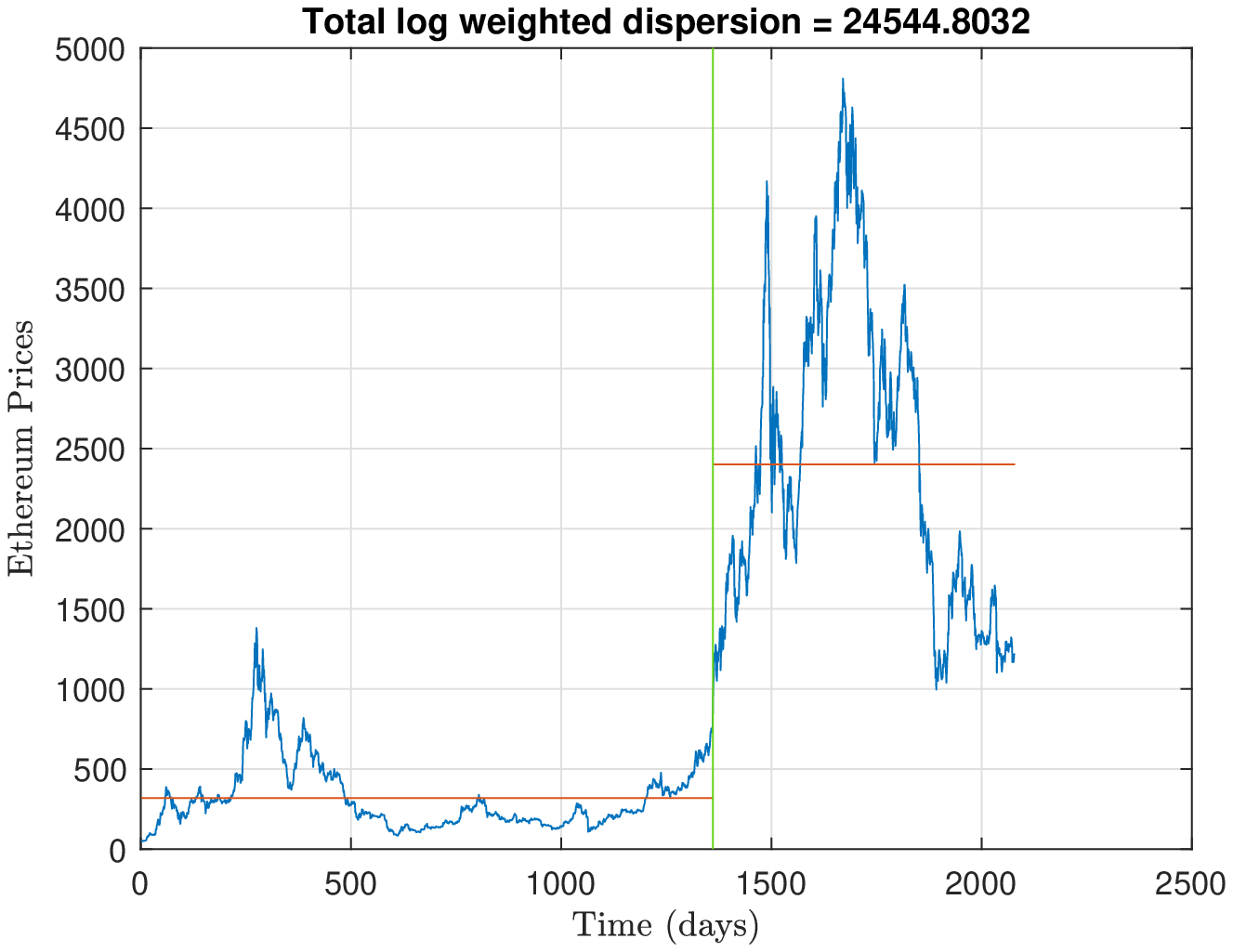}}
  \subfigure[Ethereum: Five change-points]{\label{fig:autocorr}\includegraphics[width=0.45\textwidth]{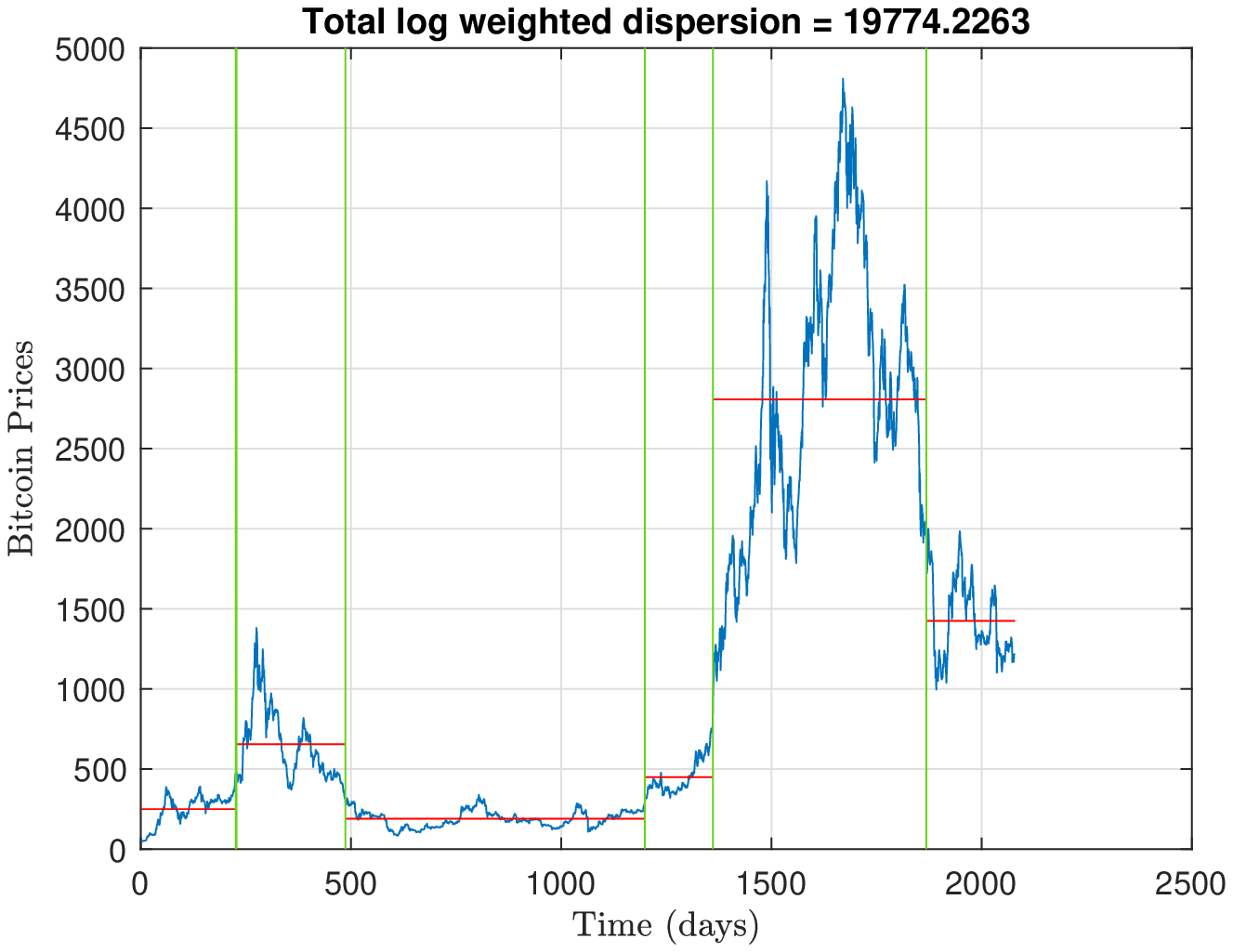}}
  \subfigure[Litecoin: One change-point]{\label{fig:log-prices}\includegraphics[width=0.45\textwidth]{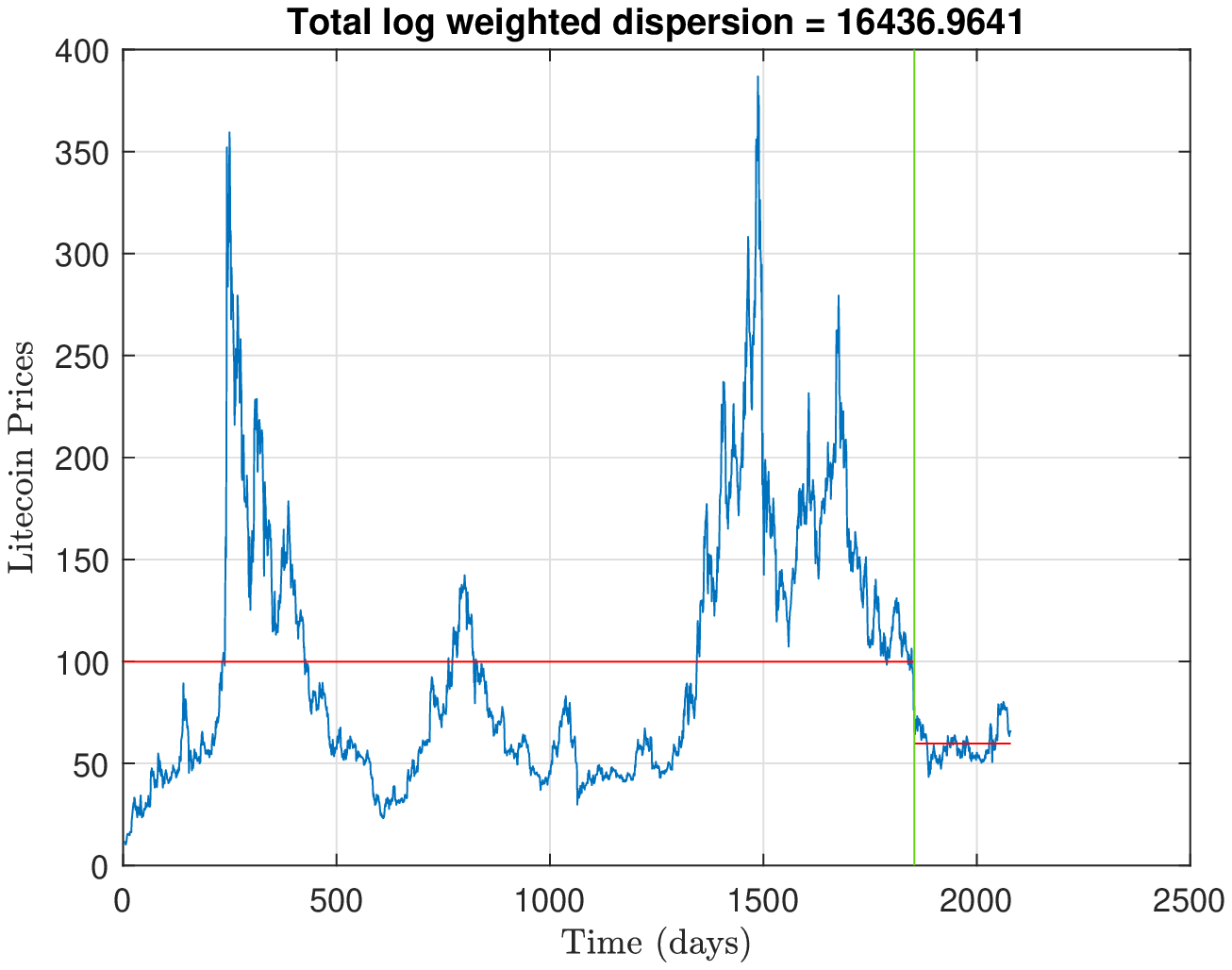}}
  \subfigure[Litecoin: Four change-points]{\label{fig:autocorr}\includegraphics[width=0.45\textwidth]{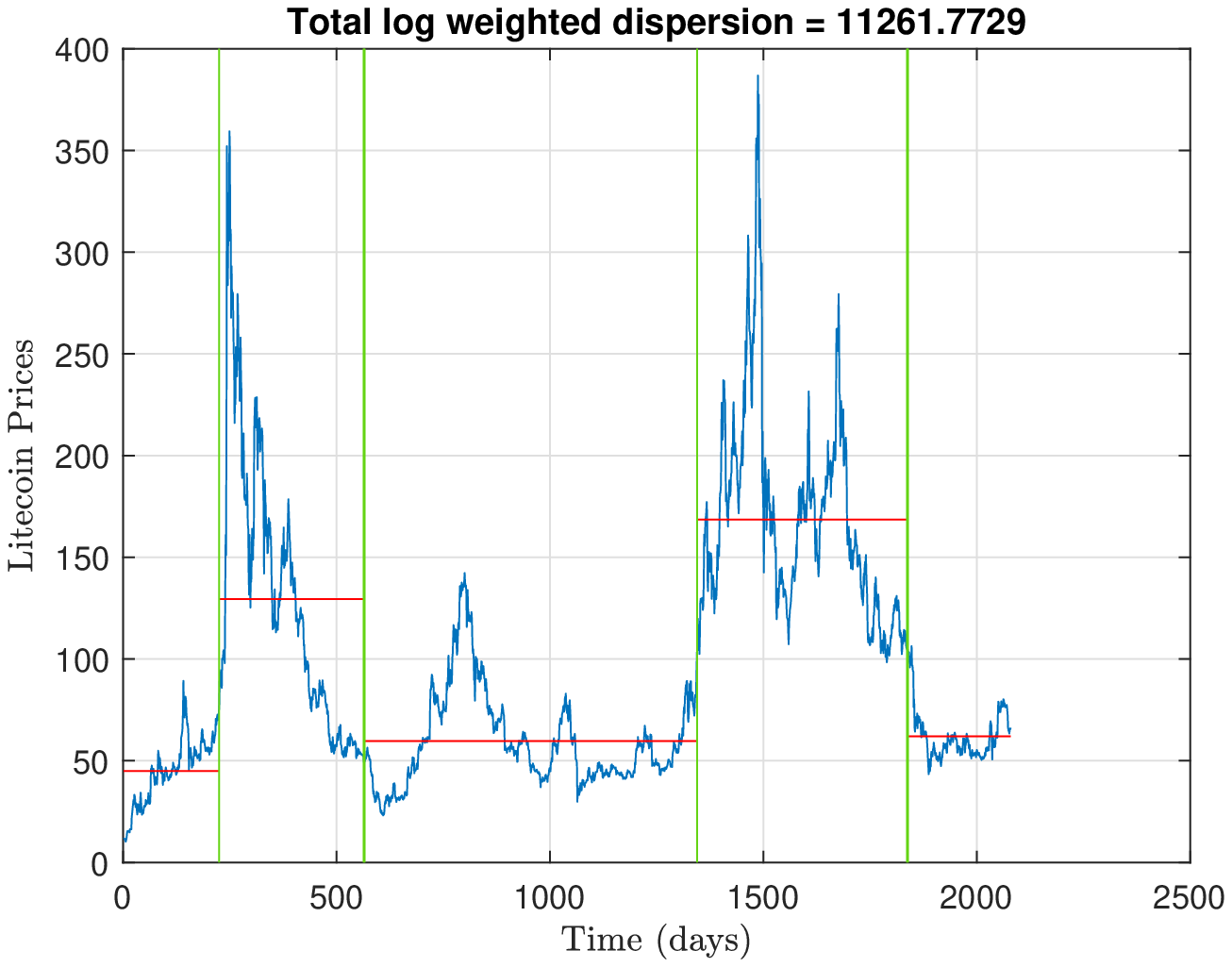}}
  \caption{Dividing the sampled period of the cryptocurrencies into a number of intervals using a test
  for multiple change-point detection (Left: the most significant change-point. Right: several change-points).}
  \label{fig:plots3}
\end{figure}

Implemented according to several methodologies like
\cite{[BER11],[ELM14],[LVL05]}, the change-point detection is a
statistical approach which allows to obtain a certain spectral
classification of the data. The division of the sampled period
into sub-intervals the point separating the periods where the
change was significant. On each of the adjacent intervals, it is
assumed that the statistical properties of the three
cryptocurrencies change significantly. As shown in the right-hand
side of Figure \ref{fig:plots3}, the classification results in
several levels. For Bitcoin, we count 4 exchange points occurring
respectively on 9/10/2017, 11/12/2019, 5/11/2020 and 13/6/2022.
For Ethereum, there were 5 change dates which were 25/11/2017,
13/8/2018, 25/6/2020, 3/1/2021 and 26/5/2022. Finally, for
Litecoin there were 4 change points detected on 24/11/2017,
29/10/2018, 18/12/2020 and 24/4/2022. In order to also know the
most significant break over the sampled period, a dichotomous
change-point detection test is applied and its results are plotted
in left-hand side of Figure \ref{fig:plots3}. The summary
statistics and long memory estimates of each cryptocurrency over
the sub-intervals separated by change-points (right-hand side of
Figure \ref{fig:plots3}) are summarized in Table
\ref{Tab:subperiods}. The tendency and dispersion results show a
significant change from one level to another. It is also clear and
obvious that an increase in trend is followed by an increase in
variation. This means that the excessive rise in prices is
characterized by a high level of risk and vice versa. Looking at
the properties of scale (long-range dependence), in particular the
GPH test and the Hurst expose, the most notable is that the prices
of the three cryptocurrencies change in type and level of
persistence. This is in fact consistent with the assumption that
returns are multi-fractional increments \citep{[BER13]}. It is now
important to study the structural multifractality of the three
cryptocurrencies.

%\begin{figure}[!ht]
%  \centering
%  \subfigure[S\&P 500]{\label{fig:log-prices}\includegraphics[width=0.45\textwidth]{Distributions}}
%  \subfigure[S\&P 500 log-returns]{\label{fig:autocorr}\includegraphics[width=0.45\textwidth]{Spectra}}
%  \caption{.}
%  \label{fig:plots2}
%\end{figure}

\begin{figure}[!ht]
  \centering
  \subfigure[Distributions]{\label{fig:log-prices}\includegraphics[width=0.45\textwidth]{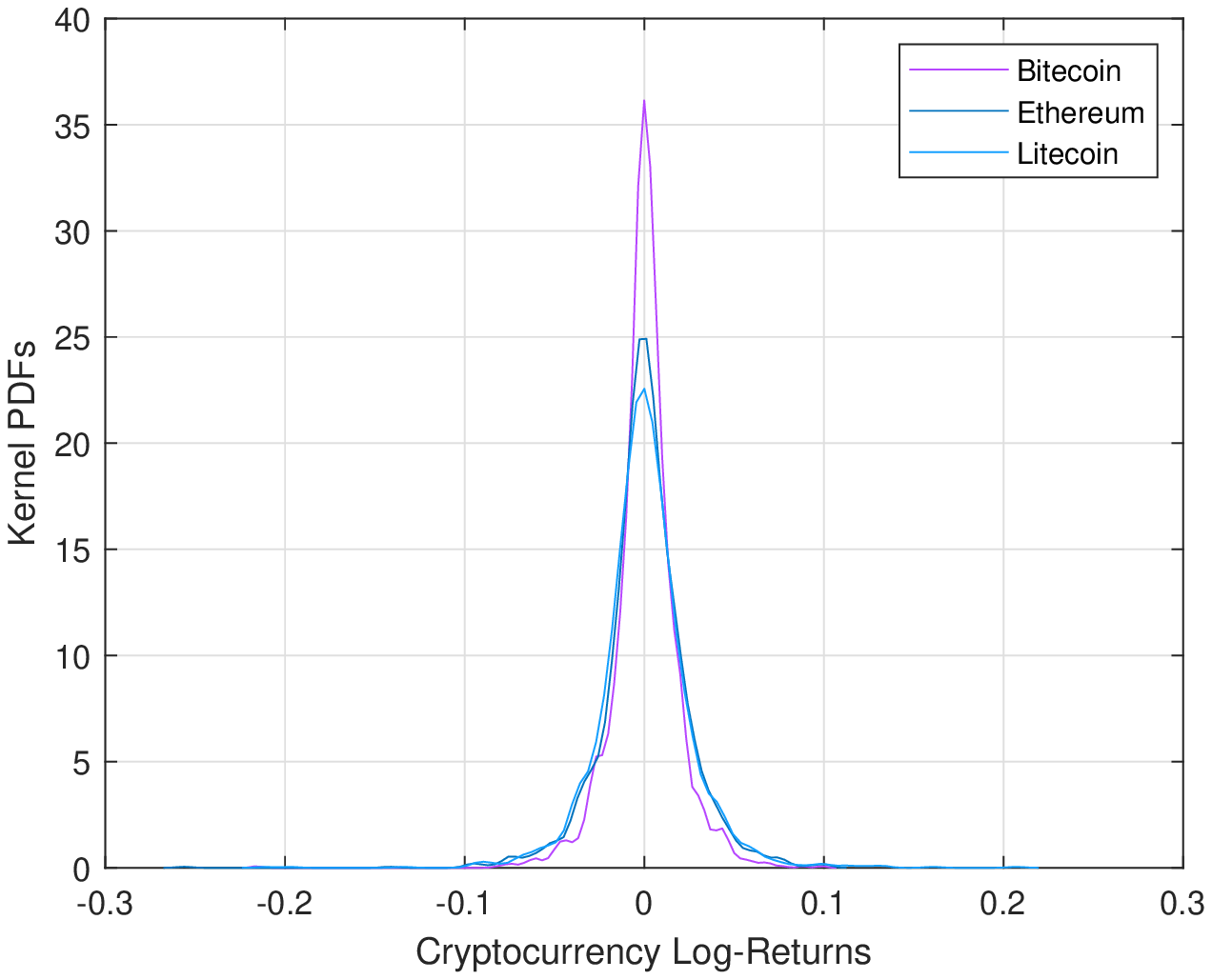}}
%  \subfigure[Bitcoin's structured multifractal spectrum]{\label{fig:log-prices}\includegraphics[width=0.45\textwidth]{Bitcoin-legendre}}
  \subfigure[Frequency spectra]{\label{fig:autocorr}\includegraphics[width=0.45\textwidth]{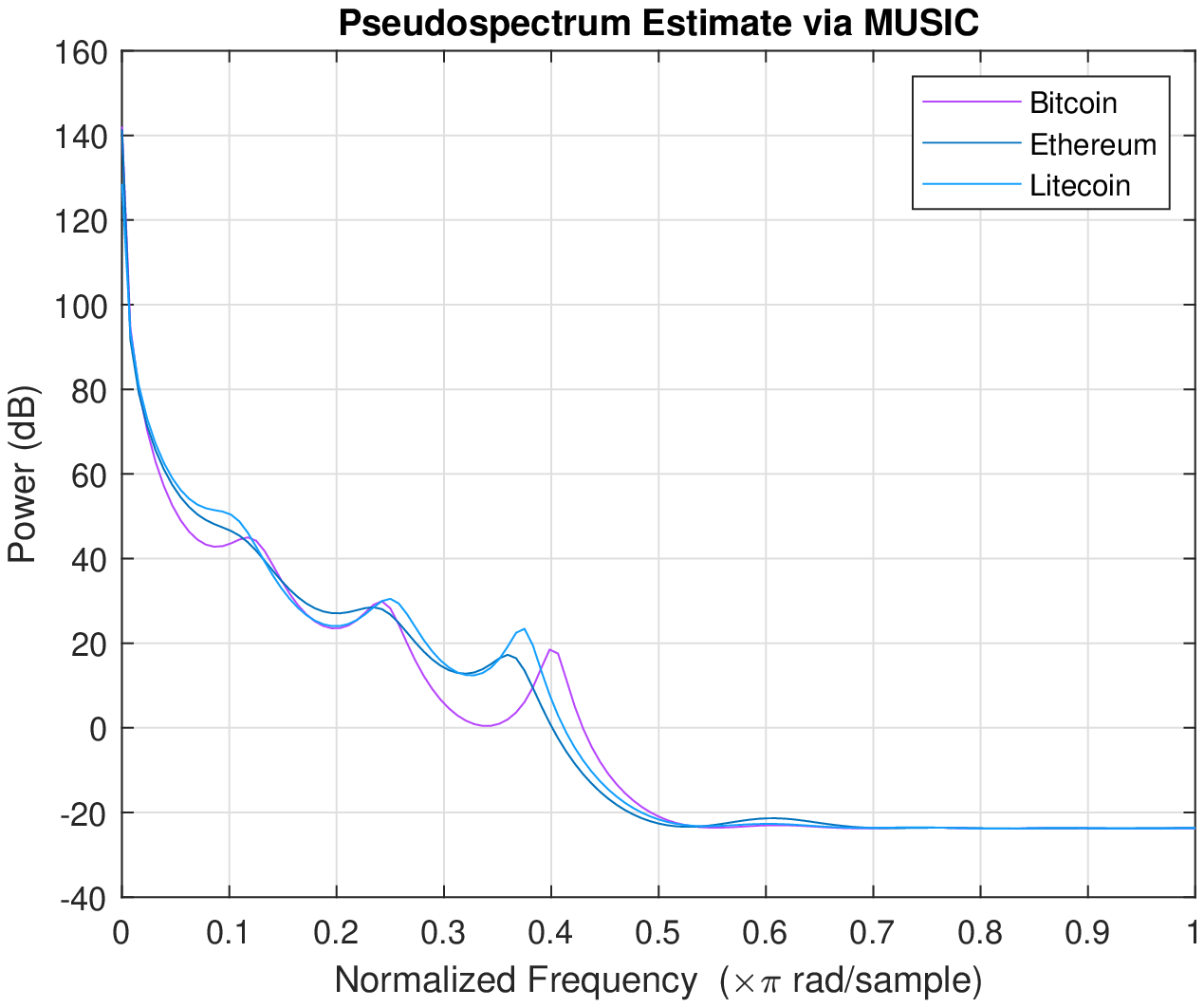}}
%  \subfigure[Ethereum's's structured multifractal spectrum]{\label{fig:log-prices}\includegraphics[width=0.45\textwidth]{Ethereum-legendre}}
%  \subfigure[Multifractal spectra]{\label{fig:autocorr}\includegraphics[width=0.45\textwidth]{MultifractalSP}}
%  \subfigure[Litecoin's structured multifractal spectrum]{\label{fig:log-prices}\includegraphics[width=0.45\textwidth]{Litecoin-legendre}}
  \caption{Distributions of densities and frequencies (MUSIC: multiple signal classification) for the three cryptocurrency prices.}
  \label{fig:plots3.}
\end{figure}

\begin{center}
\begin{table}[!ht]
\centering \caption{Statistics and tests. Estimates with asterisks
are performed on the returns (difference time series), while the
remaining ones are obtained form level data.}
\label{Tab:Descriptive}
\begin{tabular}{lcccccc}
\hline
Statistics &\qquad~~\qquad~~\qquad~~&       Bitcoin           &\qquad~~&             Ethereum             &\qquad~~&              Litecoin                         \\
\hline ~~~~~~Frequency             &&                daily             &&            daily            &&                  daily                                    \\
~~~~~~Minimum                      &&              1176.8              &&           48.50             &&                 10.21                                     \\
~~~~~~Maximum                      &&              67527.9             &&           4808.38           &&                 386.82                                    \\
~~~~~~Mean                         &&              18423               &&           1039.11           &&                 95.53                                     \\
~~~~~~Std deviation                &&              16759               &&           1160.89           &&                 63.04                                     \\
~~~~~~Coef. variation              &&              90.97\%             &&           111.72\%          &&                 66.00\%                                   \\
~~~~~~Kurtosis$^{*}$               &&              13.7444             &&           23.2517           &&                 60.7950                                   \\
~~~~~~Skewness$^{*}$               &&             -0.1825              &&          -1.0370            &&                 0.9796                                    \\
~~~~~~JB test$^{*}$                &&              10006.1             &&           35883.3           &&                 289543                                    \\
~~~~~~ADF test$^{*}$               &&             -47.1051             &&          -49.3239           &&                 -47.3131                                  \\
~~~~~~GPH$^{*}$ ($\hat{d}$)        &&              0.14261             &&           0.10744           &&                 -0.1173                                   \\
~~~~~~BDS test$^{*}$               &&              12.0452             &&           15.6776           &&                 15.6328                                   \\
 \hline
\end{tabular}
\end{table}
\end{center}

\begin{center}
\begin{table}[!ht]
\centering \caption{Distribution of outliers and extreme values.}
\label{Tab:extremes}
\begin{tabular}{lcccccc}
\hline
Statistics &\qquad~~\qquad~~\qquad~~&       Bitcoin           &\qquad~~&             Ethereum             &\qquad~~&              Litecoin                         \\
~~~~~~ $\sharp$ Low outliers       &&                 0                &&              0              &&                    0                                      \\
~~~~~~ $\sharp$ High outliers      &&                 15               &&              90             &&                    38                                     \\
~~~~~~ $\sharp$ Low extremes       &&                 0                &&              0              &&                    0                                      \\
~~~~~~ $\sharp$ High extremes      &&                 0                &&              0              &&                    1                                      \\
 \hline
\end{tabular}
\end{table}
\end{center}

\begin{center}
\begin{table}[!ht]
\centering \caption{Summary statistics and long memory estimates
of each cryptocurrency over the sub-intervals separated by
change-points. Estimates with asterisks are performed on the
returns (difference time series).}\label{Tab:subperiods}
\begin{tabular}{lcccccc}
\hline
                      &\qquad&       Bitcoin               &&               Ethereum            &&               Liitecoin                         \\
\hline
\textbf{Period 1}     &&  \emph{14 Apr 17 - 9 Oct 17}      &&    \emph{14 Apr 17 - 25 Nov 17}   &&  \emph{14 Apr 17 - 24 Nov 17}
\\\\
Mean                  &&     2903.45           &&               250.490                      &&          45.134                    \\
Std dev.              &&     1056.91           &&               93.7091                      &&          16.555                    \\
IQR                   &&     1735.15           &&               106.130                      &&          24.570                    \\
Hurst                 &&     0.22505           &&               0.93235                      &&          0.45969                   \\
GPH$^{*}$ ($\hat{d}$) &&    -0.36109           &&               0.23476                      &&          -0.12448                  \\
\hline \textbf{Period 2}     && \emph{10 Oct 17 - 11 Dec 19} &&
\emph{26 Nov 17 - 13 Aug 18 }   &&   \emph{25 Nov 17 - 29 Oct 18}
\\\\
Mean                  &&     7728.34           &&               652.211                       &&          129.43                   \\
Std dev.              &&     2908.34           &&               226.730                       &&          68.709                   \\
IQR                   &&     3236.70           &&               321.410                       &&          92.640                   \\
Hurst                 &&     0.62582           &&               0.66038                       &&          0.61674                  \\
GPH$^{*}$ ($\hat{d}$) &&     0.06515           &&              -0.14604                       &&          -0.16879                 \\
\hline \textbf{Period 3}     && \emph{12 Dec 19 - 5 Nov 20} &&
\emph{14 Aug 18 - 25 Jul 20}      &&   \emph{30 Oct 18 - 18 Dec
20}
\\\\
Mean                  &&     9395.78          &&                 188.902                      &&          59.7041                    \\
Std dev.              &&     1859.21          &&                 51.6290                      &&          23.4560                    \\
IQR                   &&     2508.85          &&                 80.6170                      &&          28.6850                    \\
Hurst                 &&     0.61835          &&                 0.52227                      &&          0.55140                    \\
GPH$^{*}$ ($\hat{d}$) &&     0.18060          &&                -0.22222                      &&          0.08100                    \\
\hline \textbf{Period 4}     && \emph{6 Nov 20 - 13 Jun 22} &&
\emph{26 Jul 20 - 3 Jan 21}       &&   \emph{19 Dec 20 - 24 Apr
22}
\\\\
Mean                  &&     42280.8           &&               446.410                       &&          168.481                    \\
Std dev.              &&     11748.4           &&               105.901                       &&          51.1713                    \\
IQR                   &&     14677.7           &&               145.810                       &&          61.0690                    \\
Hurst                 &&     0.57706           &&               0.64929                       &&          0.29227                    \\
GPH$^{*}$ ($\hat{d}$) &&    -0.03314           &&               0.10624                       &&         -0.17831                    \\
\hline \textbf{Period 5}     && \emph{14 Jun 22 - 22 Dec 22} &&
\emph{4 Jan 21 - 26 May 22}        &&   \emph{25 Apr 22 - 22 Dec
22}
\\\\
Mean                  &&     19886.5           &&                2808.45                      &&          62.1801                    \\
Std dev.              &&     2176.17           &&                885.903                      &&          12.2534                    \\
IQR                   &&     2377.25           &&                1308.28                      &&          12.5731                    \\
Hurst                 &&     0.25147           &&                0.54718                      &&          0.39737                    \\
GPH$^{*}$ ($\hat{d}$) &&     0.02084           &&                0.02785                      &&          0.05454                    \\
\hline
\textbf{Period 6}     &&       ---             && \emph{27 May 21 - 22 Dec 22}                 &&                       ---        \\\\
Mean                  &&       ---             &&       1422.94                                &&                       ---          \\
Std dev.              &&       ---             &&       239.391                                &&                       ---          \\
IQR                   &&       ---             &&       384.142                                &&                       ---          \\
Hurst                 &&       ---             &&       0.62192                                &&                       ---          \\
GPH$^{*}$ ($\hat{d}$) &&       ---             &&       0.22322                                &&                       ---          \\
\hline
\end{tabular}
\end{table}
\end{center}

\begin{figure}[!ht]
  \centering
  \subfigure[Multifractal spectra]{\label{fig:multiFract}\includegraphics[width=0.45\textwidth]{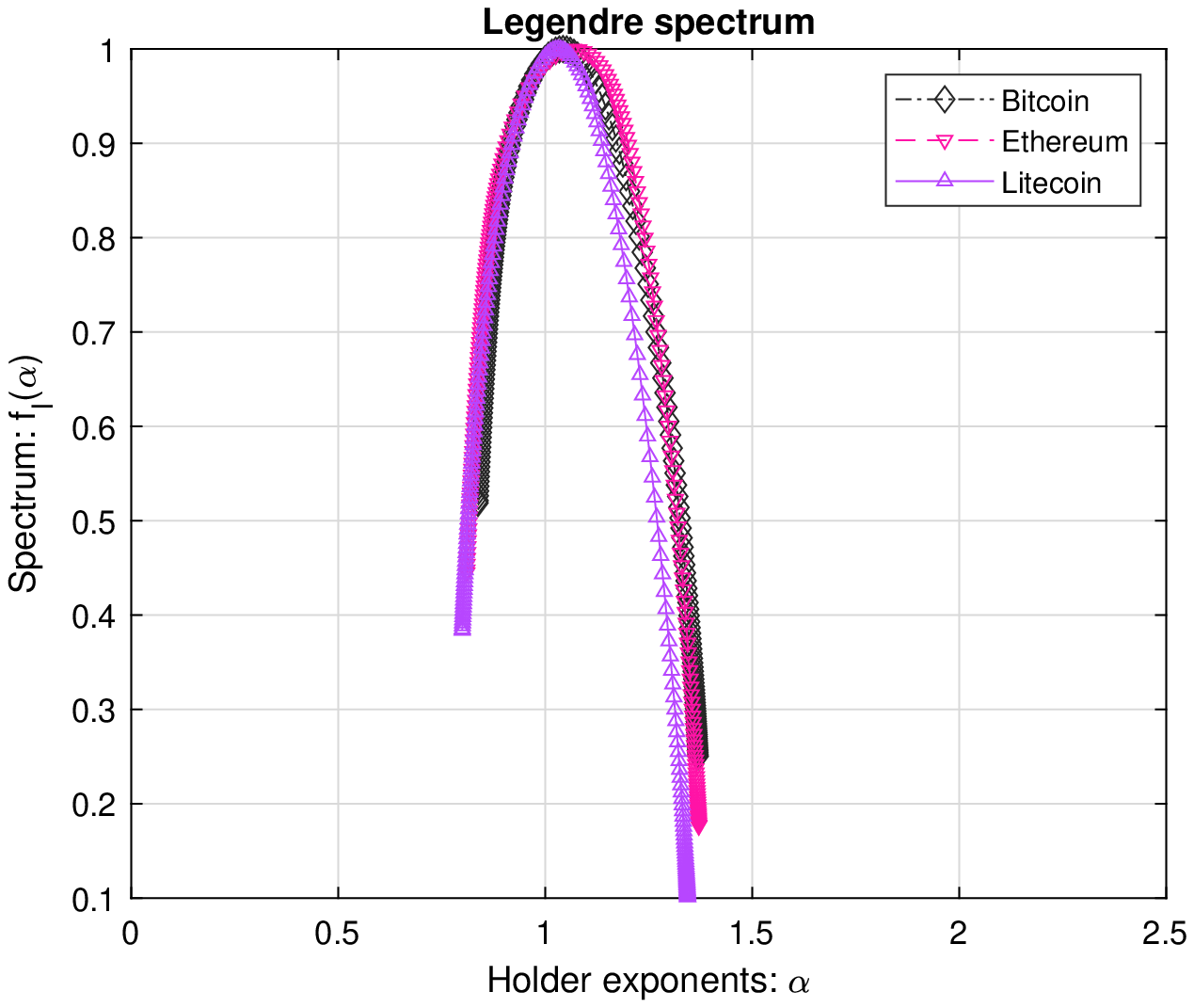}}
  \subfigure[Bitcoin's structured multifractal spectrum]{\label{fig:log-prices}\includegraphics[width=0.45\textwidth]{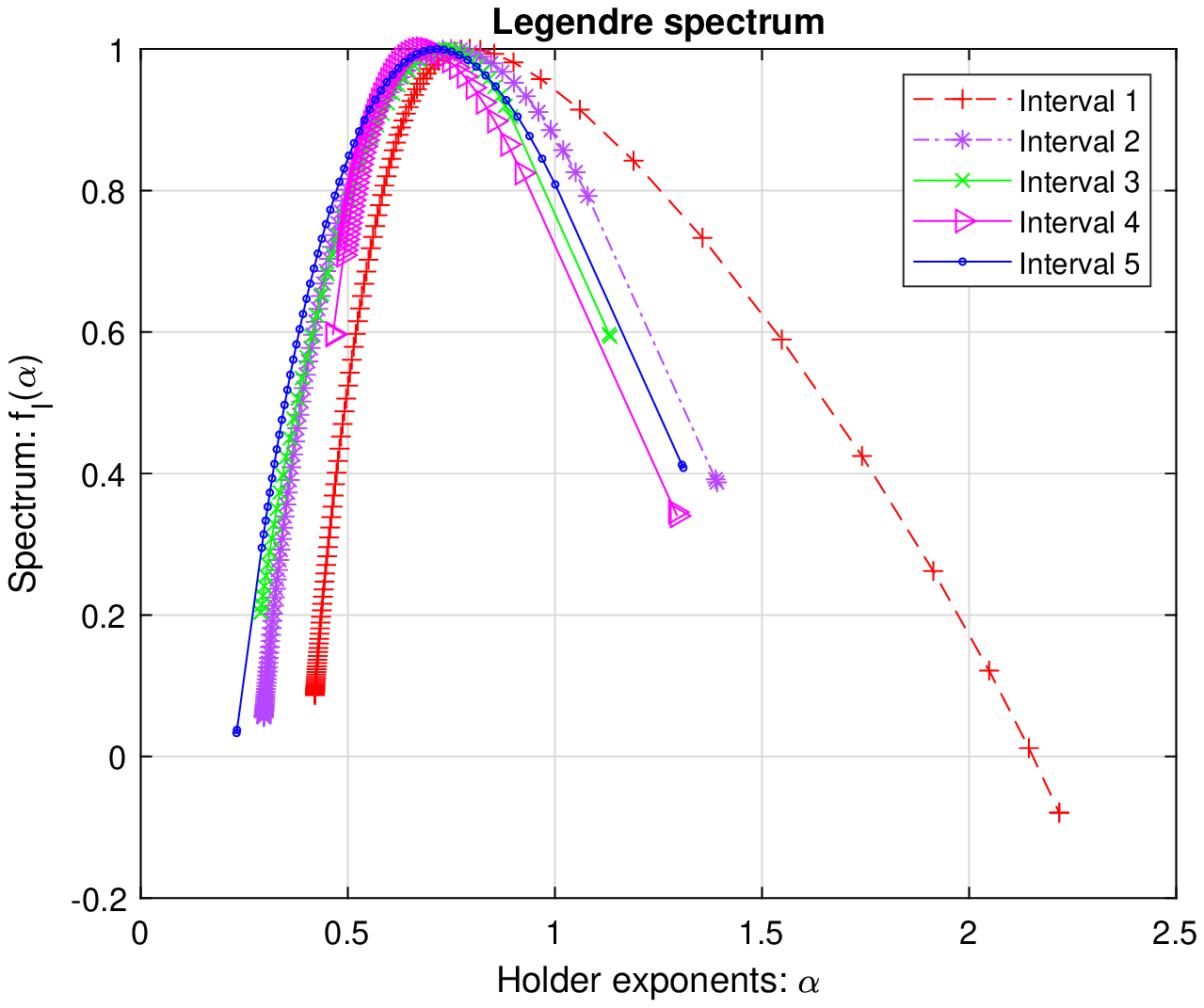}}
  \subfigure[Ethereum's's structured multifractal spectrum]{\label{fig:log-prices}\includegraphics[width=0.45\textwidth]{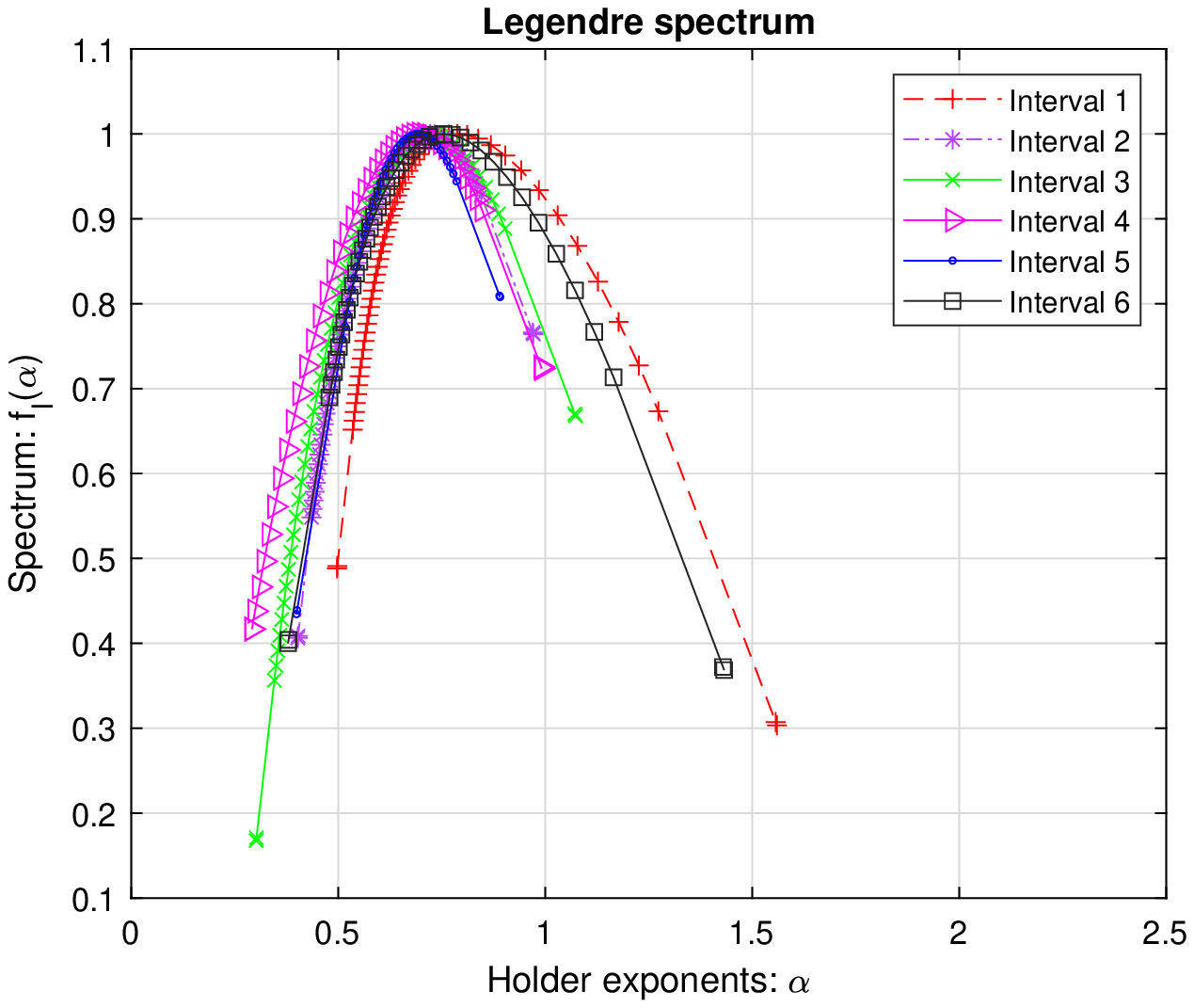}}
  \subfigure[Litecoin's structured multifractal spectrum]{\label{fig:log-prices}\includegraphics[width=0.45\textwidth]{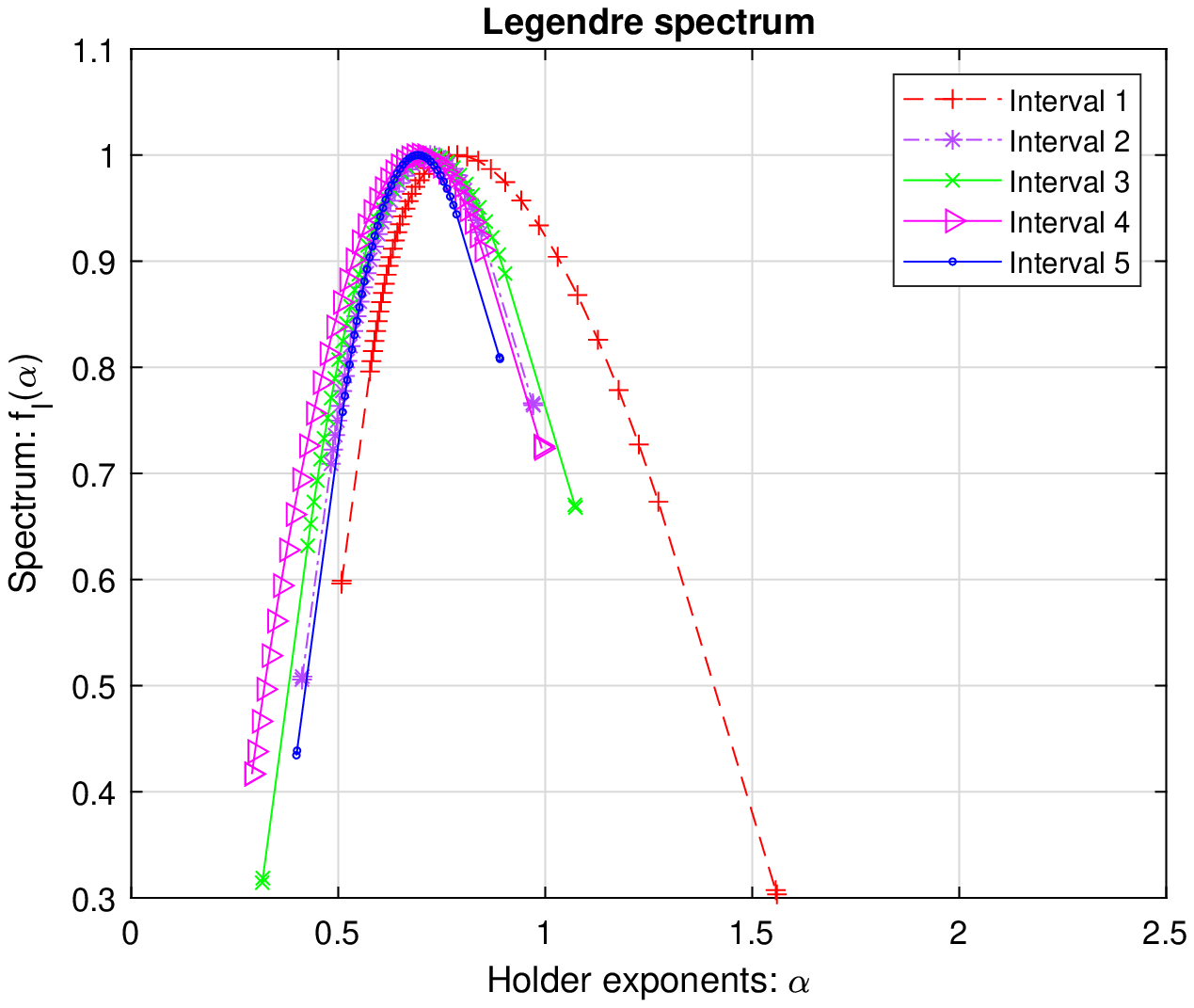}}
  \caption{Multifractality and structured multifractality of the three cryptocurrency prices.}
  \label{fig:plotsSpectM}
\end{figure}

\begin{figure}[!ht]
  \centering
  \subfigure[Bitcoin shuffled data]{\label{fig:autocorr}\includegraphics[width=0.45\textwidth]{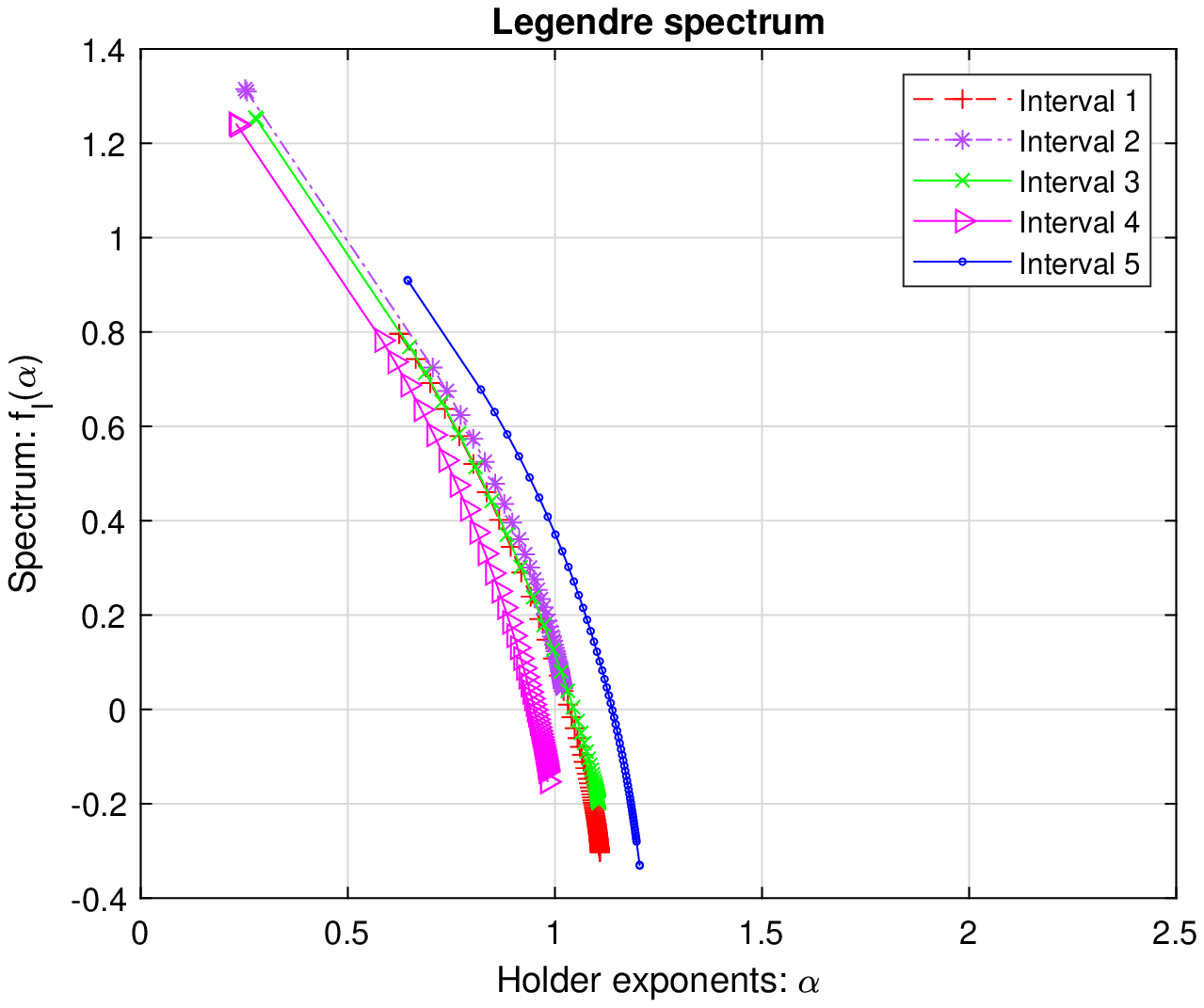}}
  \subfigure[Bitcoin surrogate data]{\label{fig:autocorr}\includegraphics[width=0.45\textwidth]{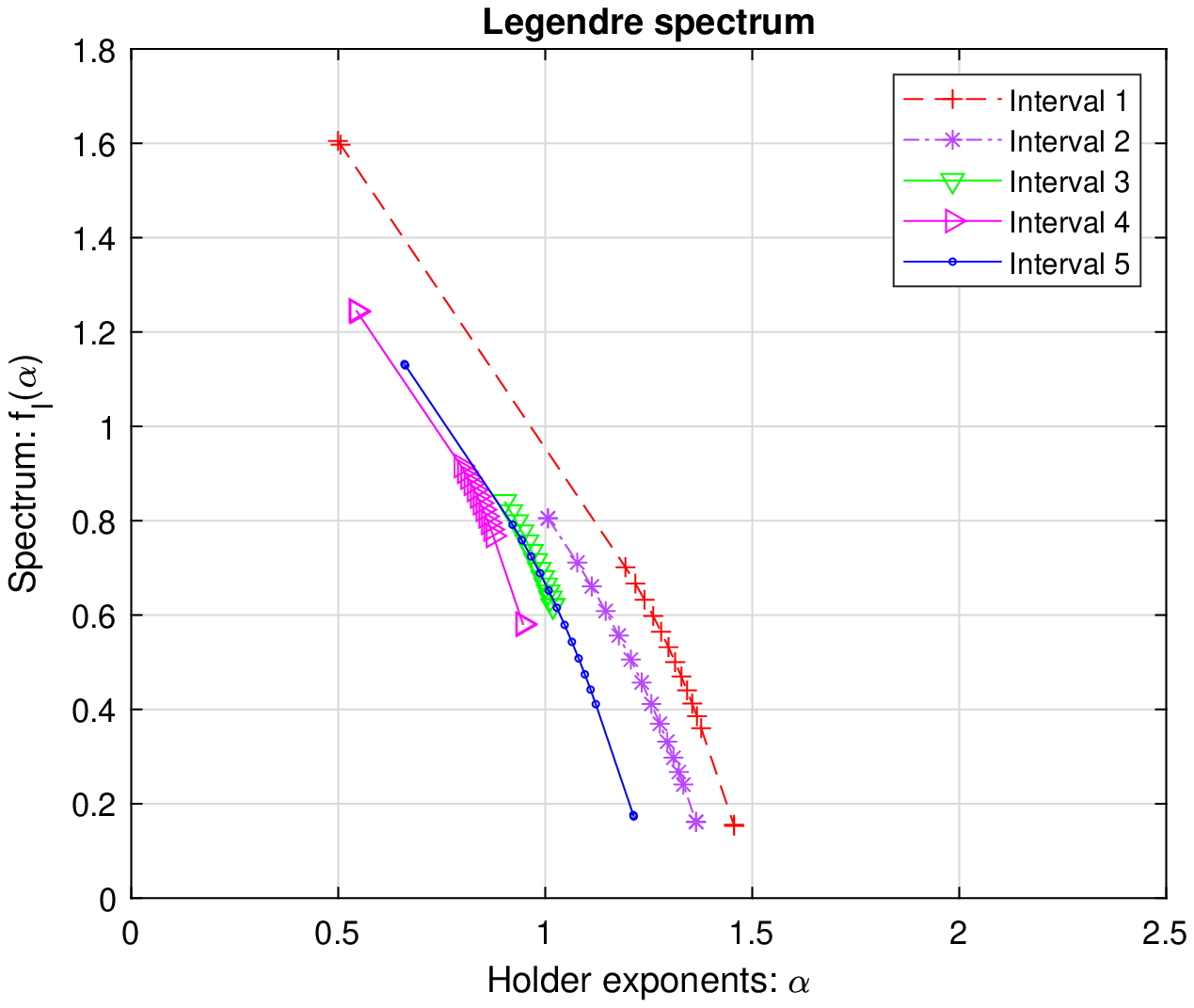}}
  \subfigure[Ethereum shuffled data]{\label{fig:autocorr}\includegraphics[width=0.45\textwidth]{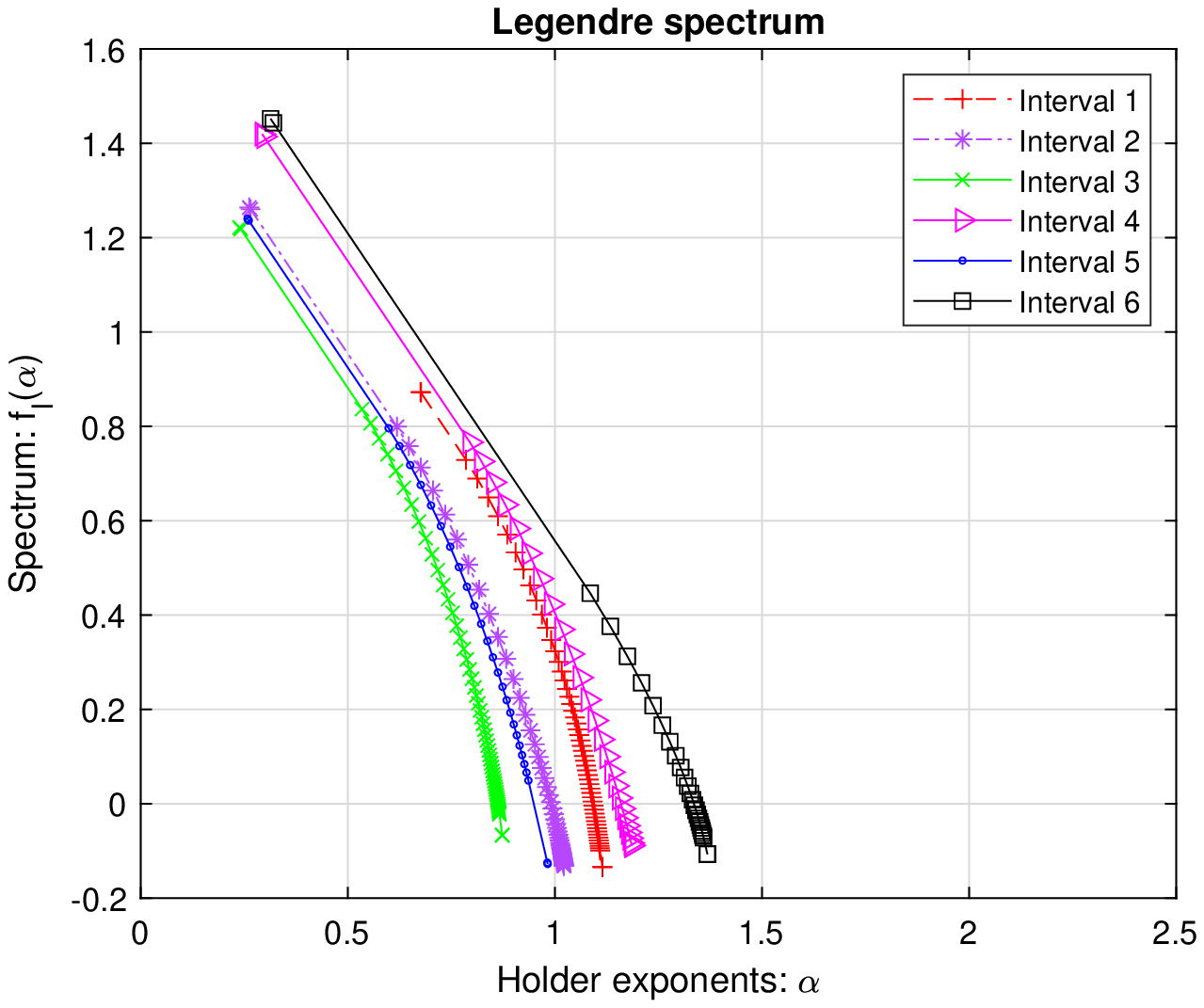}}
  \subfigure[Ethereum surrogate data]{\label{fig:autocorr}\includegraphics[width=0.45\textwidth]{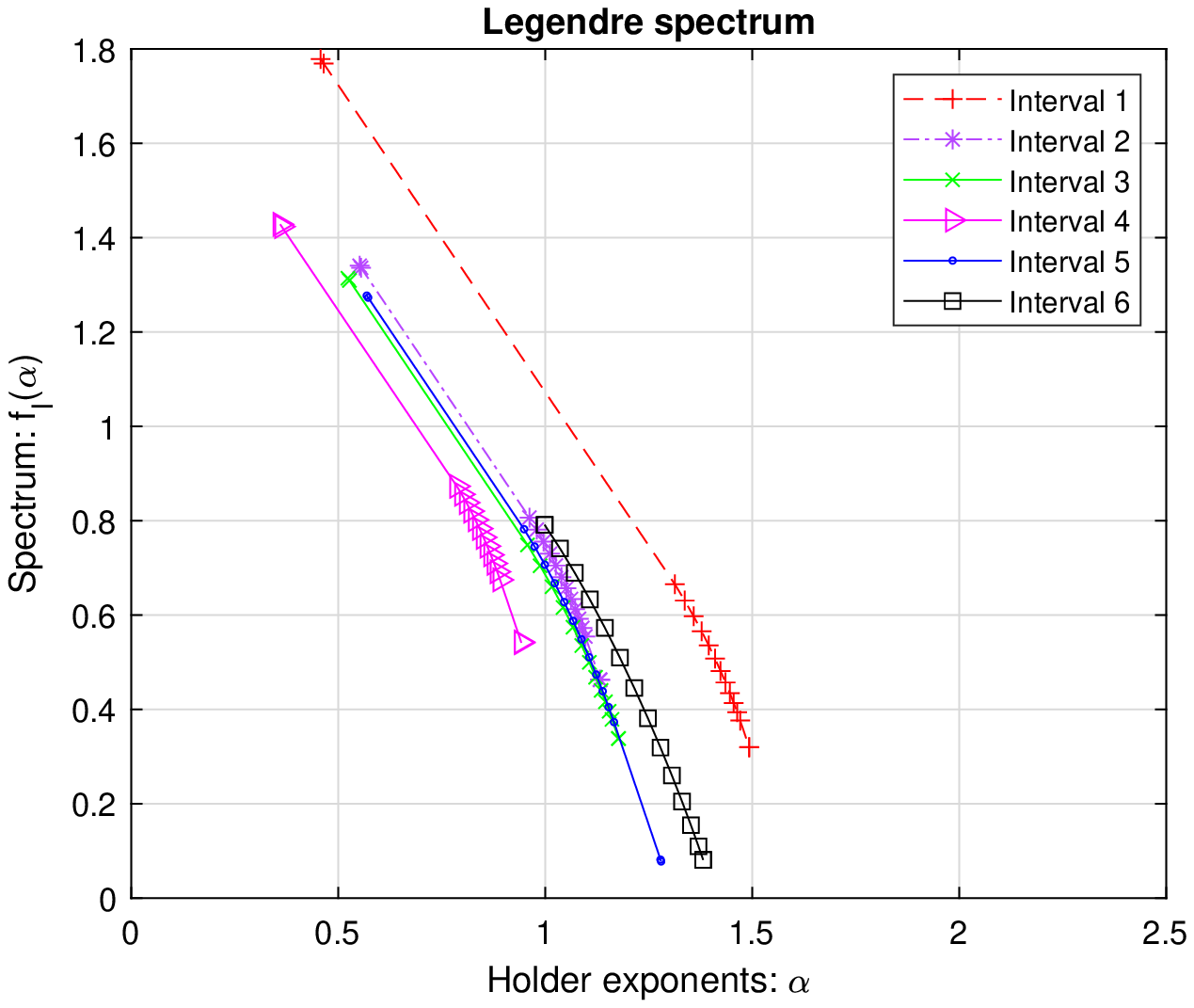}}
  \subfigure[Litecoin shuffled data]{\label{fig:autocorr}\includegraphics[width=0.45\textwidth]{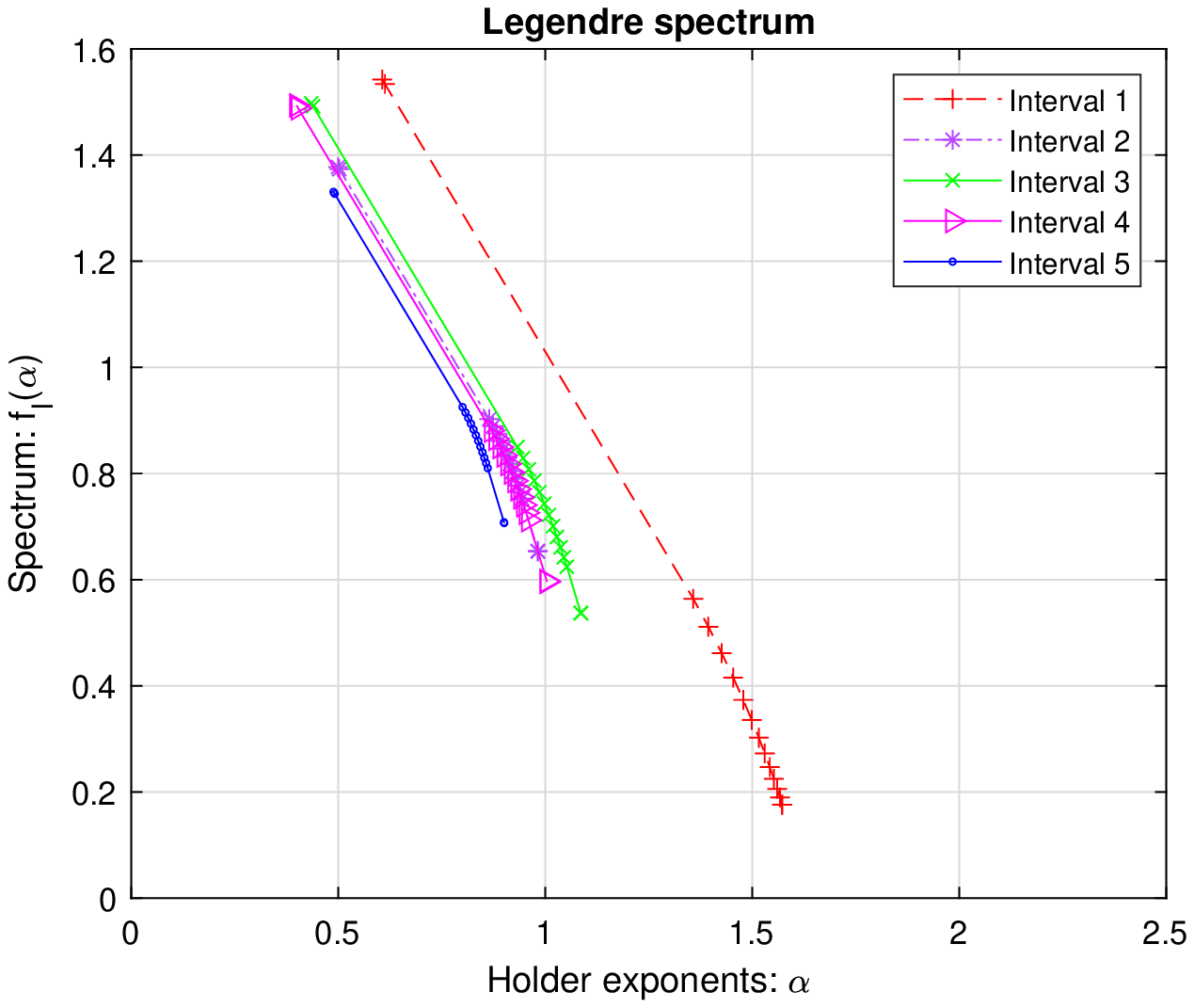}}
  \subfigure[Litecoin surrogate data]{\label{fig:autocorr}\includegraphics[width=0.45\textwidth]{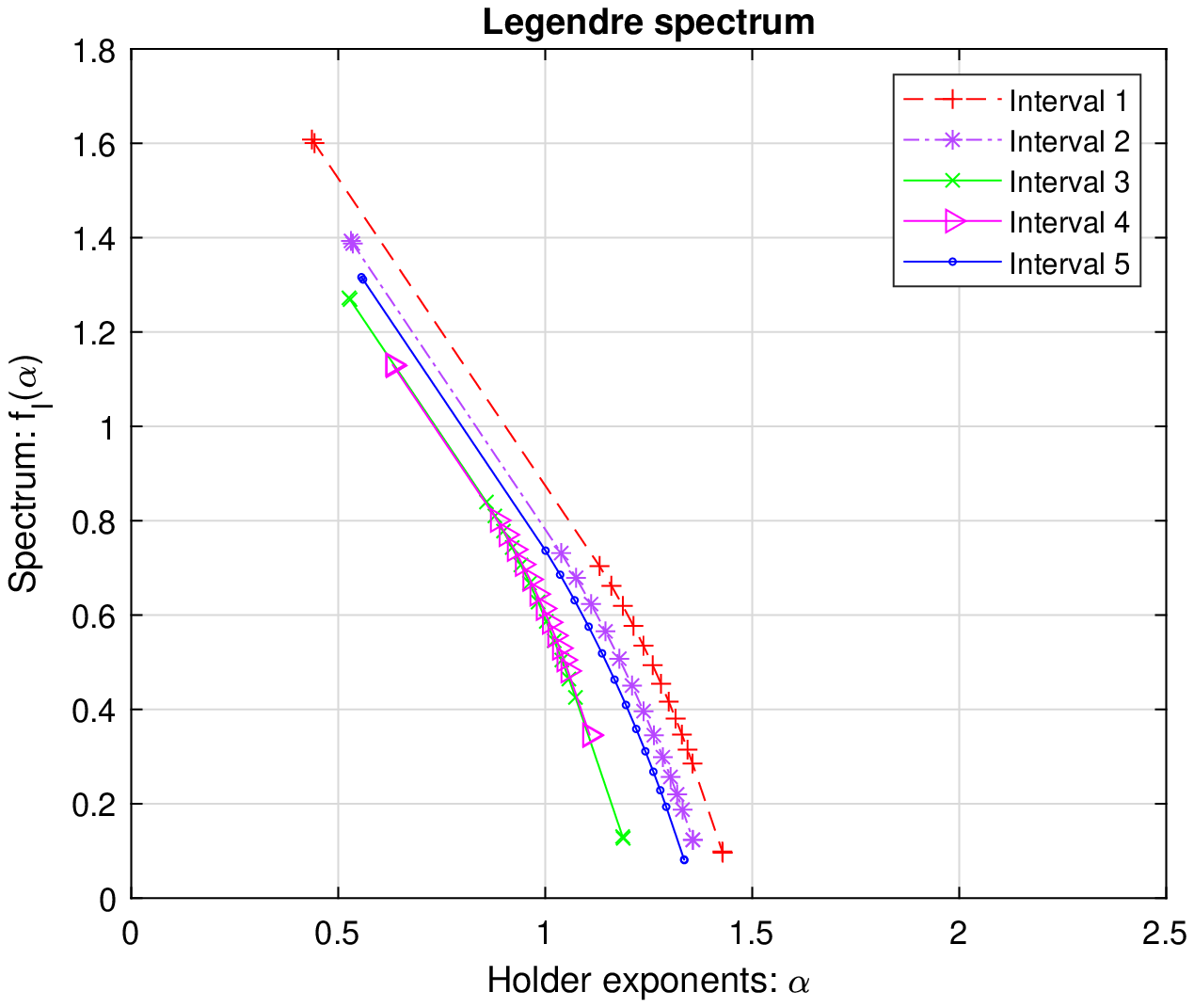}}
  \caption{Reassessing structured multifractal scaling on shuffled and surrogate cryptocurrency time series.}
  \label{fig:plotsshuffle}
\end{figure}

After the three time series are split into adjacent intervals, a
local MF-DFA is performed to each subsample. The objective is to
test structural multifractality over this tense period in the
history of the world economy. We are particularly interested in
knowing whether the efficiency of the three major cryptocurrencies
has changed significantly since the dates of the structural
breaks. To address these issues, the approach defined in section
\ref{SECT3} for testing structural multifractality is applied to
the three prices. Noteworthy that Fraclab software
\cite{[Fraclab]} or the Matlab's codes published by Ihlen
\cite{Ihlen12} can be used at this stage. The first multifractal
spectra to be interpreted in this section are those relating to
the actual data, i.e., before splitting the time series (Figure
\ref{fig:multiFract}). The most surprising in these results is
that the three cryptocurrencies have almost the same level of
multifractality because the extent of the base of the arcs
representing the spectra are almost equal. This is in addition to
the other points of resemblance underlined above, such as the
distributions of the log-returns, the frequency spectra as well as
the number of extreme values (Figure \ref{fig:plots3.}). It is now
important to show the points of dissimilarity between the three
assets since this is the question that most interests investors,
allowing them to explore alternative investment opportunities.

In Figure \ref{fig:plotsSpectM}, the local Legendre spectra show
that for the three cryptocurrencies, the local multifractality
varies significantly from one sub-period to another. This confirms
the basic hypothesis that multifractality is structural in nature,
especially in periods of successive crises. The first interval
including in particular the second and third quarters of 2017
seems to be the one where the multifractal scaling was the highest
for the three cryptocurrencies. It is also important to note that
the period of the COVID-19 outbreak characterized by record price
levels had seen the reshrinking of multifractality levels for the
three assets. It is as if the market tends to become less
inefficient in times of high risk. We finish this visual analysis
by applying the structural multifractality test on the shuffled
and surrogate data in order to verify if this multifractality of
these financial assets is to some degree related to
self-dependence, distribution, or an underlying nonlinear
correlation. Indeed, shuffling the raw series destroys the
correlations in original series and keeps the distribution of
original series, while generated surrogate series has a Gaussian
distribution and the same linear correlations as the original
series \cite{[Wu18]}. The spectra re-estimation results for each
cryptocurrency are plotted in Figure \ref{fig:plotsshuffle}. We
can clearly see that the shapes reflecting multifractality become
less obvious. This proves that the short-term dependencies, the
distribution as well as the nonlinear autocorrelations are factors
contributing significantly to this local multifractality.

%\begin{figure}[!ht]
%  \centering
%  \subfigure[S\&P 500]{\label{fig:log-prices}\includegraphics[width=0.45\textwidth]{Fig4-fractal}}
%  \subfigure[S\&P 500 log-returns]{\label{fig:autocorr}\includegraphics[width=0.45\textwidth]{Fig5-fractal}}
%  \subfigure[SSE]{\label{fig:log-prices}\includegraphics[width=0.45\textwidth]{Fig44-fractal}}
%  \subfigure[SSE log-returns]{\label{fig:autocorr}\includegraphics[width=0.45\textwidth]{Fig55-fractal}}
%  \subfigure[Nikkei 225]{\label{fig:log-prices}\includegraphics[width=0.45\textwidth]{Fig444-fractal}}
%  \subfigure[Nikkei 225 log-returns]{\label{fig:autocorr}\includegraphics[width=0.45\textwidth]{Fig555-fractal}}
%  \caption{Level and log-difference time series of S\&P 500, SSE and Nikkei 225 indexes. The samples cover a daily period
%ranging from January 1, 2018, to November 24, 2021.}
%  \label{fig:plots1}
%\end{figure}

%The q-order fluctuations $Fq(s)$ versus the scale $s$ (segment
%sample sizes) computed by MF-DFA for S\&P 500, SSE and Nikkei 225
%indexes before (left) and after (right) the change points.

\subsection{Prediction Experiments}\label{Sub-SECT3}

\begin{figure}[!ht]
  \centering
  \subfigure[Bitcoin: Period 1]{\label{fig:autocorr}\includegraphics[width=0.3\textwidth]{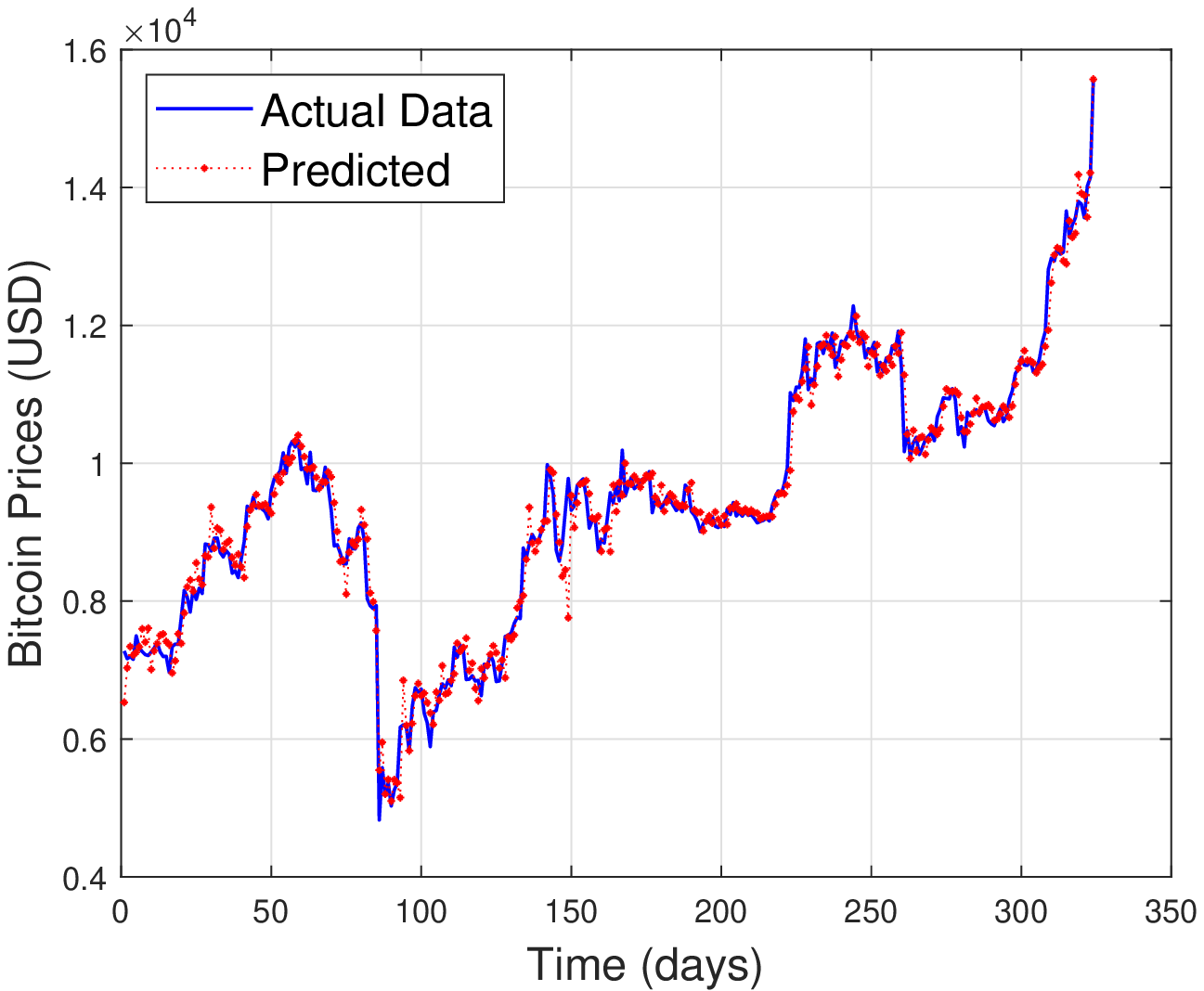}}
  \subfigure[Ethereum: Period 1]{\label{fig:autocorr}\includegraphics[width=0.3\textwidth]{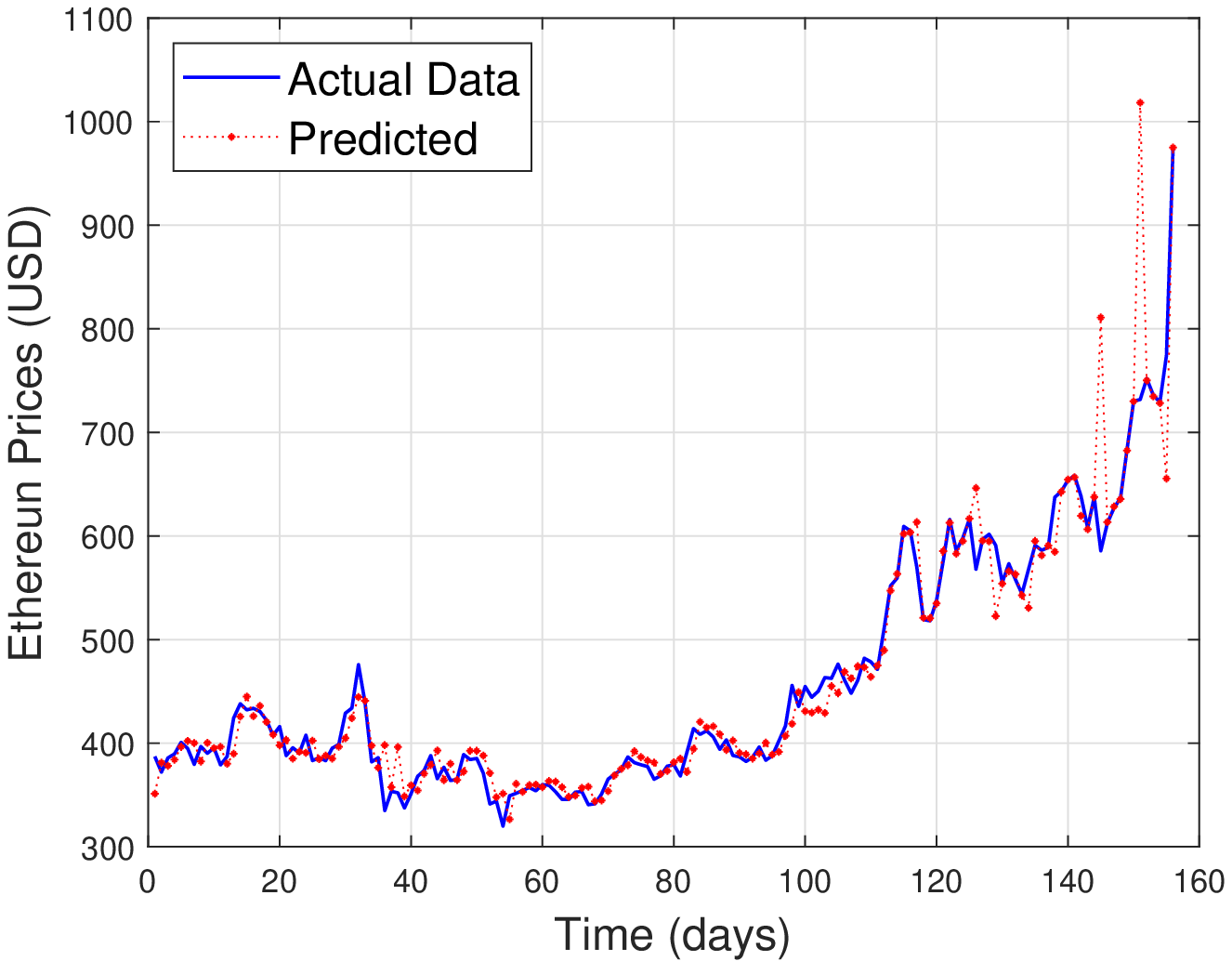}}
  \subfigure[Litecoin: Period 1]{\label{fig:autocorr}\includegraphics[width=0.3\textwidth]{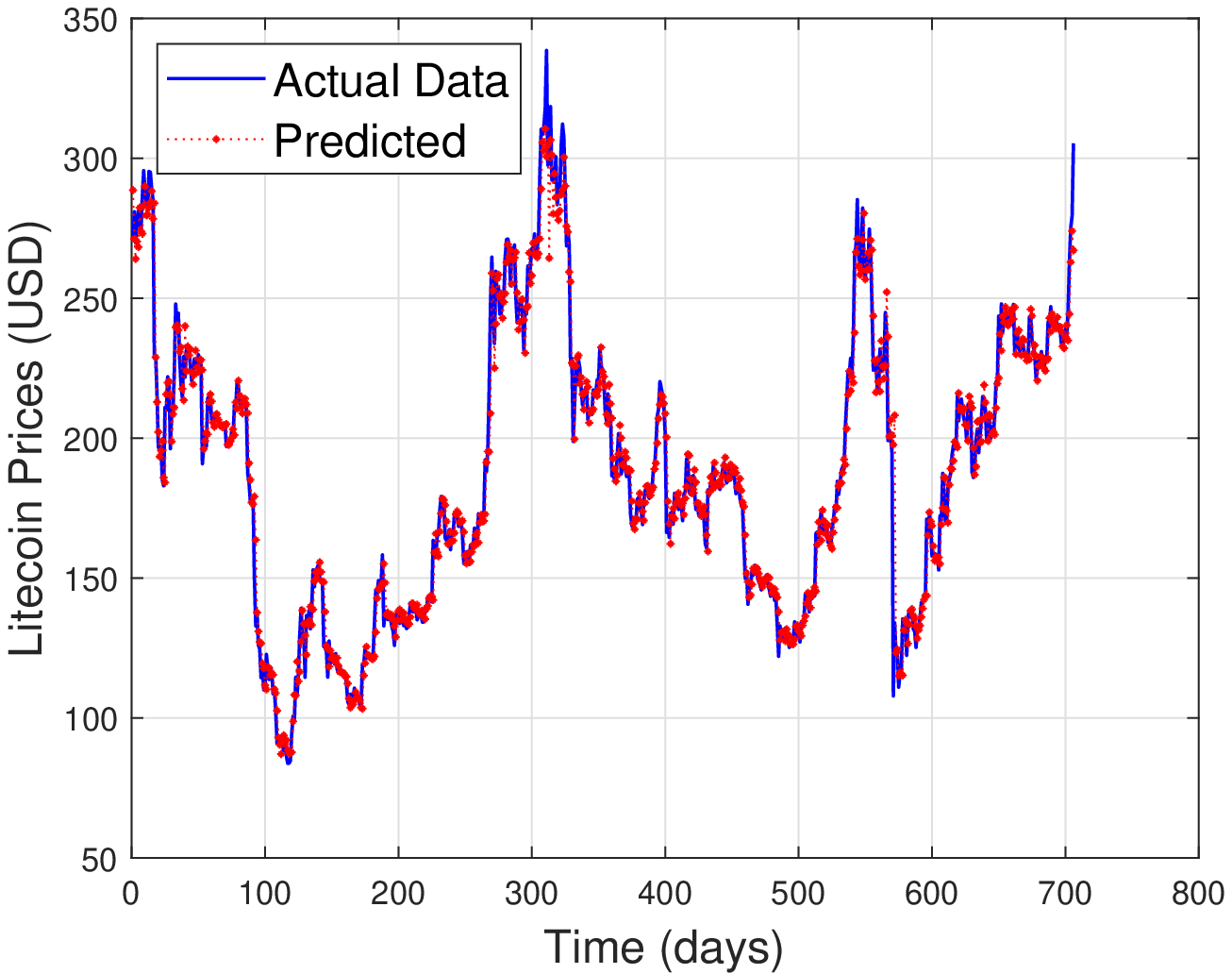}}
  \subfigure[Bitcoin: Period 2]{\label{fig:autocorr}\includegraphics[width=0.3\textwidth]{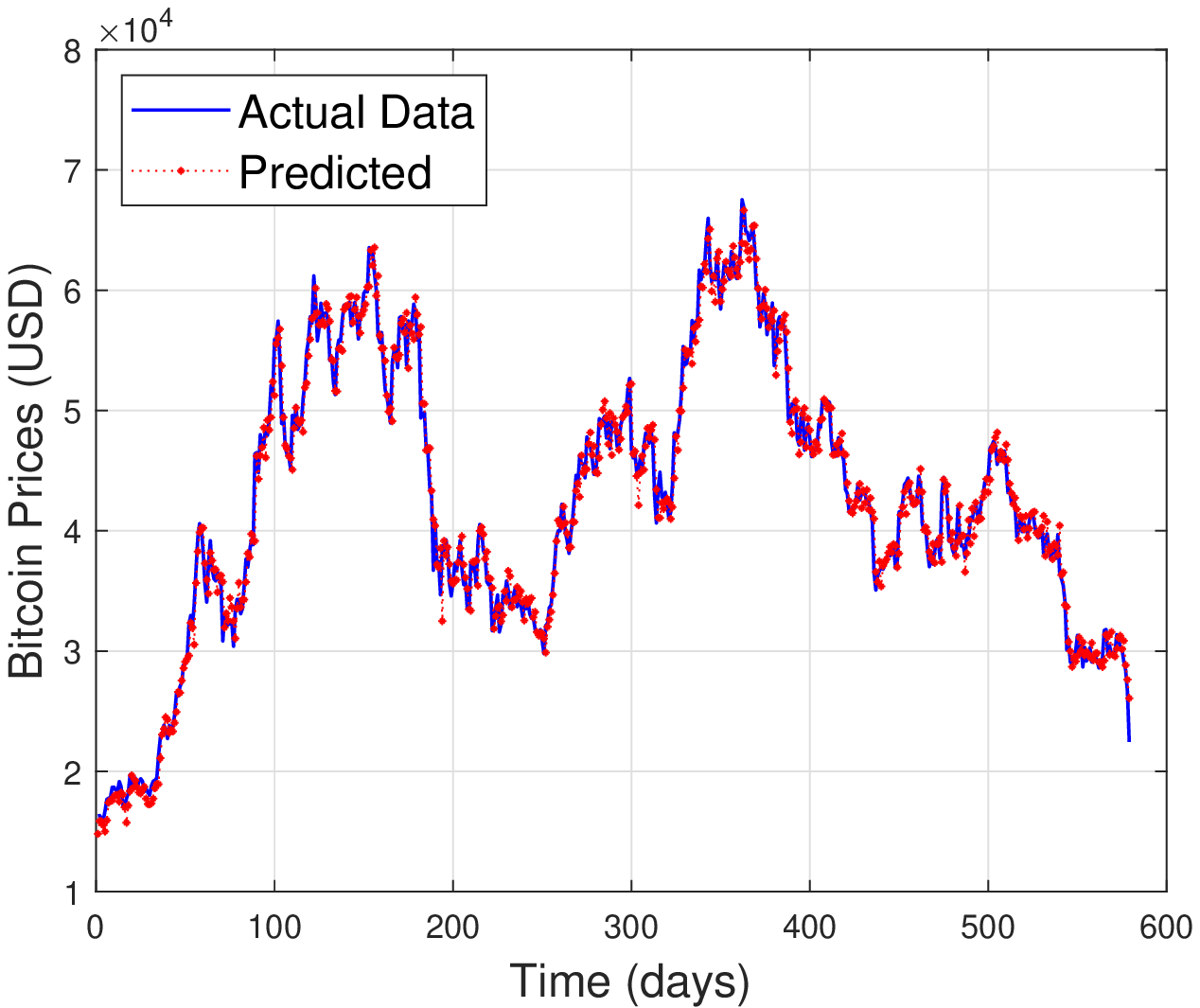}}
  \subfigure[Ethereum: Period 2]{\label{fig:autocorr}\includegraphics[width=0.3\textwidth]{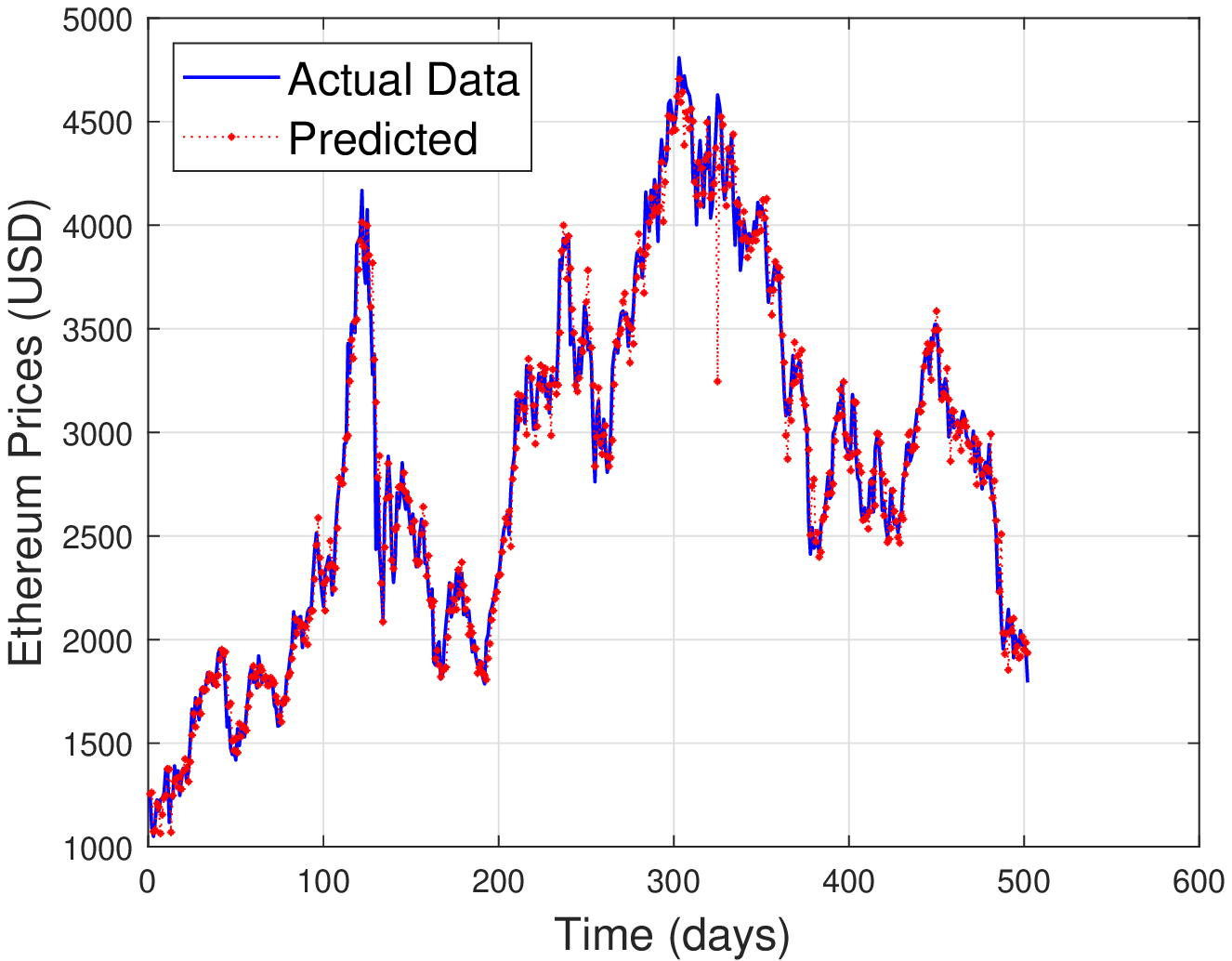}}
  \subfigure[Litecoin: Period 2]{\label{fig:autocorr}\includegraphics[width=0.3\textwidth]{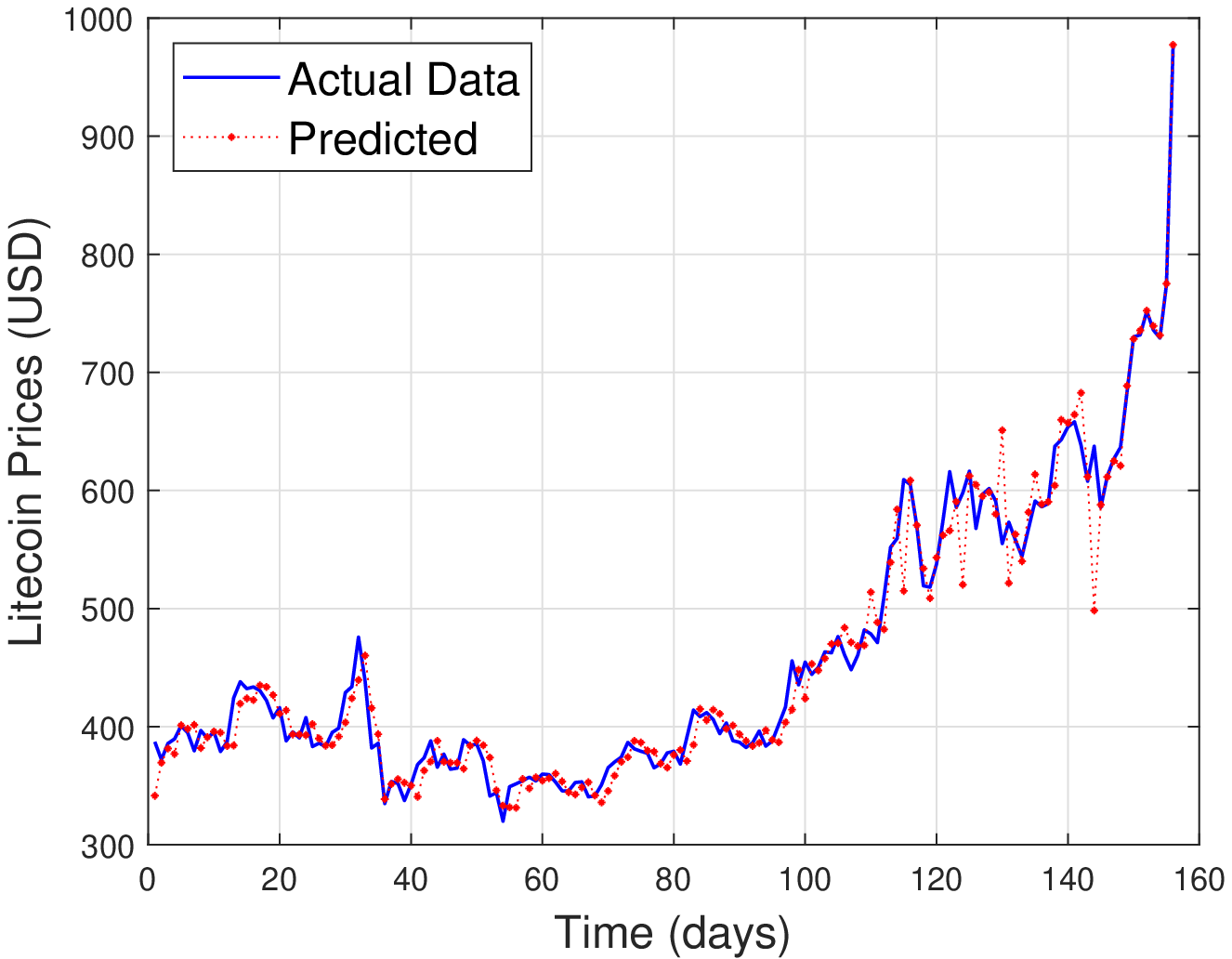}}
  \subfigure[Bitcoin: Period 3]{\label{fig:autocorr}\includegraphics[width=0.3\textwidth]{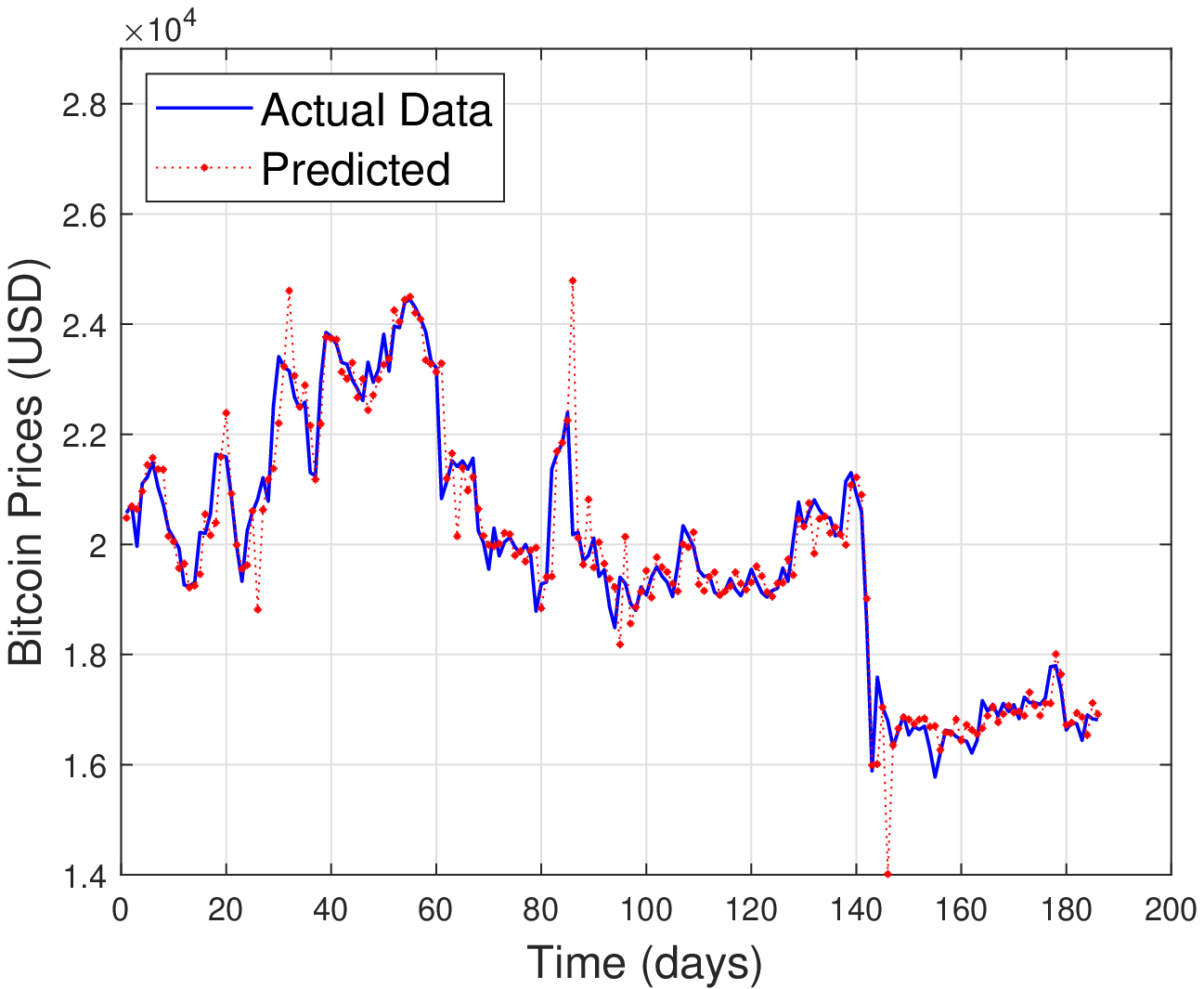}}
  \subfigure[Ethereum: Period 3]{\label{fig:autocorr}\includegraphics[width=0.3\textwidth]{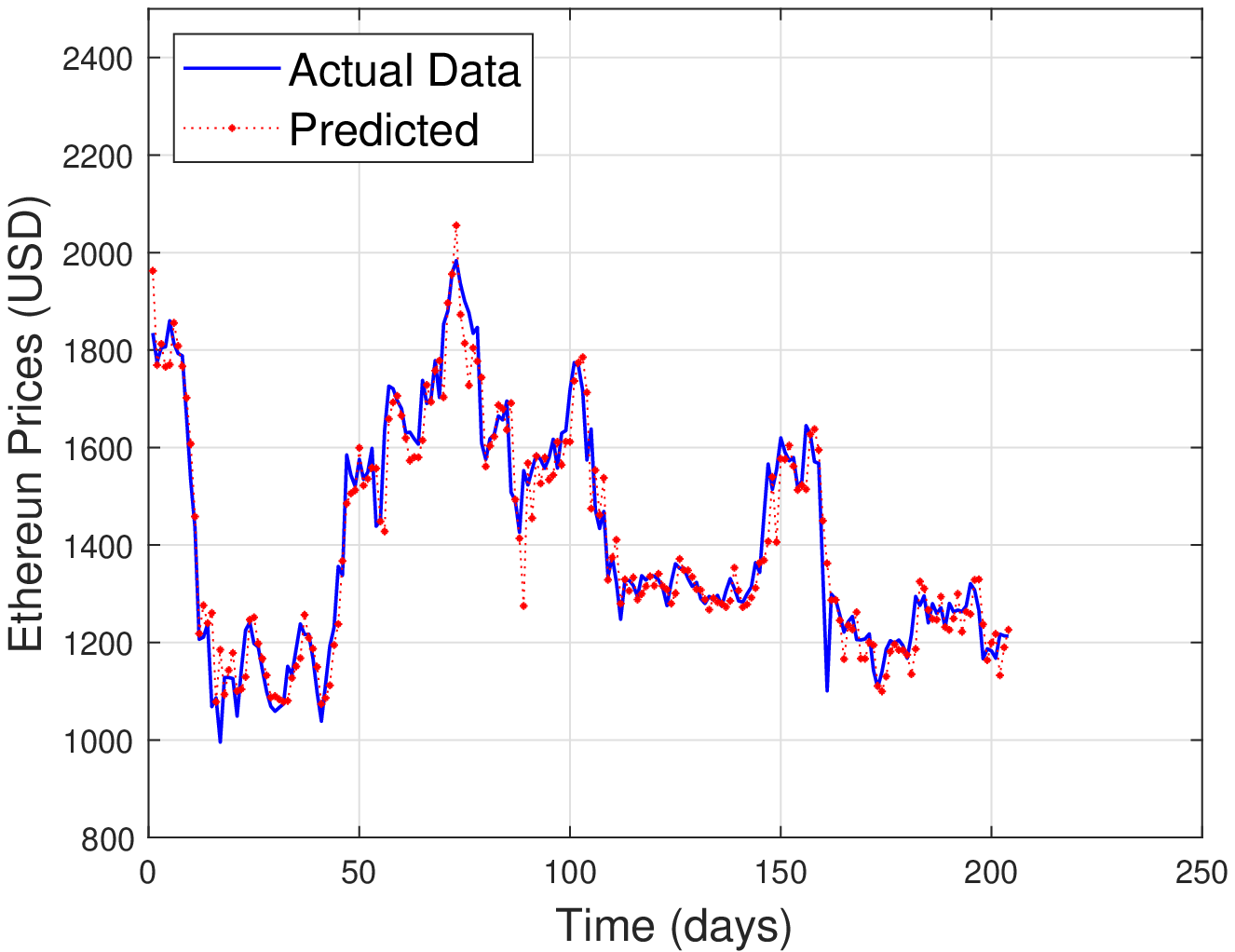}}
  \subfigure[Litecoin: Period 3]{\label{fig:autocorr}\includegraphics[width=0.3\textwidth]{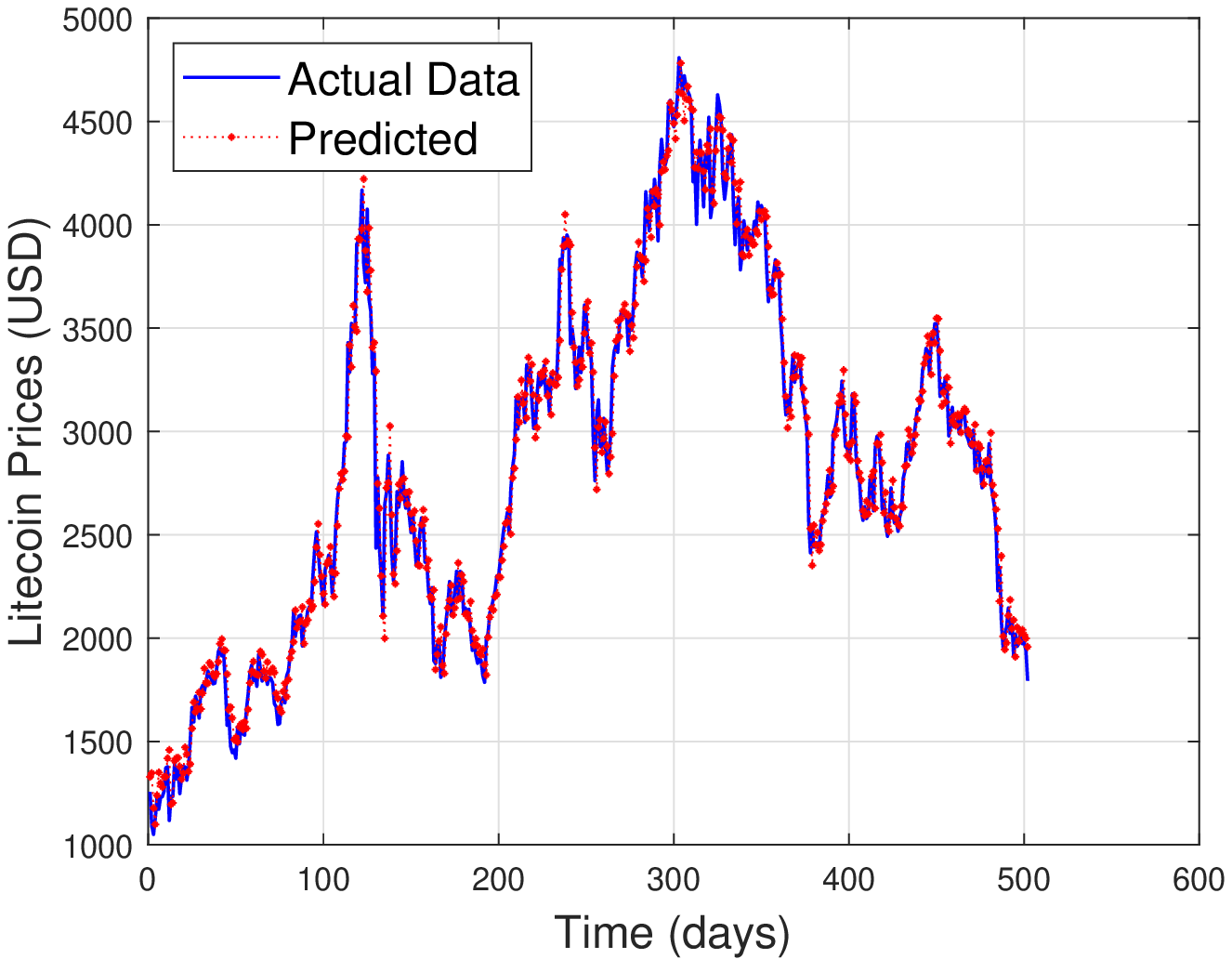}}
  \caption{Prediction results of the three cryptocurrency prices over the last three intervals (see Table \ref{Tab:subperiods}).}
  \label{fig:forecasts}
\end{figure}

We finish our experiments in this paper with prediction tests
applied to the three cryptocurrencies on each of the periods fixed
above. In particular, we compare two types of fractional
differentiation preceding learning using a neural network of the
backpropagation category. This methodology remains one of the most
widely used for forecasting and decision-making \cite{[Tang21]}.
The objective is to compare the efficiency of a structured
fractional modeling with that of a simple fractional modeling in a
pure machine learning framework. For the first case, the
fractional differentiation parameters by intervals are as
estimated using the GPH method in Table \ref{Tab:subperiods}. As
for the overall differentiation, this is done based on the single
parameter given in Table \ref{Tab:Descriptive}, also according to
the GPH's methodology. Noteworthy that Velasco's \cite{[Vela99]}
fractional differentiation technique and Shimotsu's Matlab
code\footnote{\url{https://shimotsu.web.fc2.com/styled-3/}.} are
used in this step. The results of the predictions are summarized
in Table \ref{Tab:mapeRes}. The mean absolute percentage error
(MAPE) is used to express the prediction accuracy as a ratio
defined by the formula:
\begin{equation}\label{eqn:mape}
{\displaystyle {\mbox{MAPE}}={\frac
{1}{n}}\sum_{t=1}^{n}\left|{\frac
{A_{t}-F_{t}}{A_{t}}}\right|\times 100\%},
\end{equation}
where $A_{t}$ is the actual value and $F_{t}$ is the forecast
value. The results of the errors of this prediction phase are
reported in Table \ref{Tab:mapeRes}. The predictions made here are
complete reconstructions of the series of the three prices on each
of the sub-periods all based on a Neural Autoregressive (NAR)
model with 20 hidden layers, with sigmoid type activations and
$p=5$ no lags. A Levenberg-Marquardt algorithm is used to train
this neural network. We can note a relative improvement in the
accuracy of predictions when differentiations are adapted to
breakpoints occurring, which confirms that the locally
multifractal time series require local transformations in order to
obtain better performance from autoregressive models. In Figure
\ref{fig:forecasts}, as a graphical illustration, we plot the
predictions made by the NAR model based on the multi-fractional
integration over the last three time periods. It is clear that the
predictions reach an acceptable level of accuracy. It is now
important to extend this type of neural model to equip it with a
causal decision-making loop allowing it to also integrate other
endogenous factors \cite{[Saad14]}.

\begin{center}
\begin{table}[!ht]
\centering \caption{Summary statistics and long memory estimates
of each cryptocurrency over the sub-intervals separated by
change-points. Estimates with asterisks are performed on the
returns (difference time series).} \label{Tab:mapeRes}
\begin{tabular}{lccccccc}
\hline
                         & Method  &\qquad\qquad&       ~~Bitcoin~~               &\qquad&       Ethereum      &\qquad&    Liitecoin \\
\hline
\textbf{Period 1}\qquad\qquad\qquad\qquad        &&&                              &&                                            &&                                    \\
                         &FD-NAR &&     237.20            &&               268.55                      &&          282.64                    \\
                         &LFD-NAR&&     234.45            &&               241.32                      &&          281.01                    \\
\hline
\textbf{Period 2}        &&&                              &&                                           &&                                    \\
                         &FD-NAR&&      337.42            &&               296.97                      &&          278.35                    \\
                         &LFD-NAR&&     145.40            &&               279.44                      &&          222.78                    \\
\hline
\textbf{Period 3}        &&&                              &&                                            &&                                    \\
                         &FD-NAR&&       189.19           &&               416.85                       &&          374.47                    \\
                         &LFD-NAR&&     180.22            &&               254.09                       &&          223.97                    \\
\hline
\textbf{Period 4}        &&&                              &&                                            &&                                    \\
                         &FD-NAR&&     288.19             &&               423.44                       &&          255.81                    \\
                         &LFD-NAR&&    284.88             &&               377.17                       &&          242.65                    \\
\hline
\textbf{Period 5}        &&&                              &&                                            &&                                    \\
                         &FD-NAR&&     303.74             &&               200.38                       &&          191.34                    \\
                         &LFD-NAR&&    192.33             &&               195.34                       &&          182.34                    \\
\hline
\textbf{Period 6}        &&&                              &&                                            &&                                    \\
                         &FD-NAR&&      --                &&               393.89                       &&          --                        \\
                         &LFD-NAR&&     --                &&                277.56                      &&          --                        \\
\hline
\end{tabular}
\end{table}
\end{center}

This study offered a new framework for multifractal data
analytics, using a data mining methodology. The structural
character of price fluctuation scaling could provide a basis for
founding a new family of financial market forecasting models
allowing decision makers to better anticipate future returns in
situations of alternating economic and investment conditions
between adverse and favorable. In particular, the new models could
better identify time series characteristics such as trend,
seasonality, and volatility, which can be particularly useful for
companies that use time series to plan and assess financial
performance. It is important to note that multifractal models are
still relatively new and little used in the practice of finance.
The results of these models are often dependent on the assumptions
and data used, so they may not always be reliable for decision
making. It is therefore important to use them in combination with
other methods and to be aware of their limitations. It would also
be interesting to extend the break detection methodology of our
approach by adapting instead a more sophisticated method like that
of the filtered derivative or the filtered derivative at p-value
\cite{[BER11],[ELM14]}, which generally assigns less importance to
fake financial bubbles.

\section{Conclusion}

In this paper; we have defined a new notion of so-called
structural multifractality allowing us to generalize the
conventional principle of scaling. We have also developed an
algorithmic approach to test this generalized multifractal
character for a given time series. The main requirements of this
approach is that it must be conducted on data sets of sufficient
size, which makes it particularly suitable for big data. We then
performed this on daily data from the three major cryptocurrencies
over a significant period in the history of the global economy.
This very particular global situation has made the irregularity of
the fluctuations of these financial assets stronger than usual,
which was moreover expected given the succession of important
events such as Brexit, COVID-19 and the Russian-Ukrainian
conflict. It was therefore essential to extend the principle of
conventional multifractality towards a more general aspect in
order to give more flexibility to the various econometric models
intended for forecasting. The mechanism for measuring structural
multifractal scaling essentially relies on a test for detecting
structural changes on the studied time series, before an MF-DFA is
performed on each of the resulting fluctuation regimes.

Visualizing the statistical results for the three cryptocurrencies
allows us to confirm the basic hypothesis that the returns of
these cryptocurrencies behave similarly to a structural
multifractal process rather than a simple multifractal process. It
seems, in fact, that the multiple breakpoints detected for each of
the time series, sometimes coinciding with the major events of the
period. These also seem to have an impact on the fluctuation
dynamics, and therefore on the efficiency of the decentralized
money market. Faced with this intermittent multifractal behavior,
investors must equip themselves with an adaptive decision
strategy, evolving with the fluctuating regime of the economic
situation. In other words, any occurrence of a significant event
on the economic scene must be a sign for investors of a change in
the valuation of financial assets. New explainable machine
learning forecasting models taking into account this generalized
multifractality are therefore developed and tested on the
different cryptocurrency prices.

In future research, it would be important to try understanding
whether exogenous factors could also have contributed to this
structural multifractality. The development and application of a
structural multifractal cross-correlation analysis including other
factors such as energy prices, the main stock market indices, or
monetary exchange rates, can for example be a step in this
direction. Multifractal cross-correlation of cryptocurrencies each
other would also be relevant since it allows to locate the
intervals of alternative investment opportunities. Finally, it
would also be interesting to conduct forecasting experiments by
designing intermittent multifractal autoregressive models as well.
Such models could be implemented in massively parallel
environments and tested on several types of economic and financial
data, such as stock markets, electricity prices, oil prices, food
prices, etc. All these perspectives can be realized in a
politico-economic environment which remains tense with the
continuation of the Russian-Ukrainian war.

%An accurate electricity price forecasts represents an advantage
%for market players facing competition.

%Despite various existing time series methods, research and
%experiences aiming to improve electricity prices forecasting
%accuracy have never stopped. In this same direction, this work
%demonstrates the strong performance of the class of multiscaled
%neural networks for handling bi- or multi-variate time series,
%especially when they are preceded by a scale-by-scale causality
%test.

%Since the deregulation of electricity markets, predicting
%electricity prices has become a fundamental tool for effectively
%managing energy systems. In fact, price forecasting is a key
%information in today's commodity markets. Especially in power
%markets, investors are still making extensive use of forecasting
%techniques either to bid or hedge against volatility.

%Indeed, in addition to its importance for profitability analysis
%and power planning, it allows authorities to better set up
%appropriate economic regulations of the services provided by
%monopoly transmission and distribution networks.

\section*{Acknowledgment}
This work was supported by the Deanship of Scientific Research
(DSR), King Abdulaziz University, Jeddah, under grant No.
(DF-000-000-0000). The authors, therefore, gratefully acknowledge
DSR technical and financial support.

\section*{Author statements}

\subsection*{Ethical approval} Not required because the study
did not touch on ethical issues requiring individual consent.

%\subsection*{Funding} This work has not received any specific grant from funding
%agencies in the public, commercial or not-for-profit sectors.

\subsection*{Competing interests} The authors declare no conflicts of interest.

\subsection*{Data availability statement} Data available on request from the authors.

%\section*{References}

\end{document}